\documentclass[prb,aps,twocolumn,showpacs,preprintnumbers,floats,amsmath,amssymb]{revtex4}

\usepackage{graphicx}
\usepackage{dcolumn}
\usepackage{bm}

\def \be{\begin{equation}}
\def \ee{\end{equation}}
\def\Hdot{H_{\text{grain}}}

\def\hc{{\it h.c.}}
\def\dst{\hat \rho}

\def\moy#1{\left\langle #1 \right\rangle}

\def\Re{\hbox{Re}}

\def\parent#1{\left(#1\right)}

\def\ket#1{\left\vert #1 \right\rangle}
\def\bra#1{\left\langle #1 \right\vert}

\def\parent#1{\left(#1\right)}

\def\Bigparent#1{\Bigl (#1\Bigr)}
\def\biggparent#1{\biggl (#1\biggr)}
\def\biggbra#1{\biggl [#1\biggr]}
\def\Sm{{\cal  S}_{m}}

\def\couplingbath{g_{B}}

\def\nanoLLG{NMD }

\begin{document}

\title{Theory of Spin Torque in a nanomagnet}

\author{O. Parcollet }
\affiliation{ Service de Physique Th{\'e}orique,  
CEA/DSM/SPhT - CNRS/SPM/URA 2306,
 CEA Saclay, 91191 Gif-sur-Yvette, France  }

\author{X. Waintal}
\affiliation{Service de Physique de l'{\'E}tat Condens{\'e},
CEA Saclay, 
91191 Gif-sur-Yvette cedex, France }

\date{\today}

\begin{abstract}
We present a complete theory of the spin torque phenomena in a ultrasmall nanomagnet coupled to
non-collinear ferromagnetic electrodes through tunnelling junctions. This model system can be described by
a simple microscopic model which captures many physical effects characteristic of spintronics:
tunneling magneto resistance, 
intrinsic and transport induced magnetic relaxation, current induced magnetization
reversal and spin accumulation. Treating on the same footing the magnetic and transport degrees of freedom,
we arrive at a closed equation for the time evolution of the magnetization. This equation is very close
to the Landau-Lifshitz-Gilbert equation used in spin valves structures. We discuss how the presence of the
Coulomb blockade phenomena and the discretization of the one-body spectrum gives some additional 
features to the current induced spin torque. Depending on the regime, the dynamic induced by the
coupling to electrode can be viewed either as a spin torque or as a relaxation process.
In addition to the possibility of stabilizing uniform spin precession states, we find that
the system is highly hysteretic: up to three different magnetic states
can be simultaneously stable in one region of the parameter space
(magnetic field and bias voltage).We also discuss how the magneto-resistance can be used to provide
additional information on the non-equilibrium peaks present in the nanomagnet spectroscopy experiments.
\end{abstract}

\maketitle

The observation of the current induced magnetic reversal\cite{katine2000} in 
Ferromagnetic-Normal metal-Ferromagnetic (FNF) spin valves has been a real breakthrough 
in the field of spintronic. Indeed the spin torque effect~\cite{slonczewski1996} at the 
root of this magnetic reversal allows one to control the magnetic configuration using
electronic current only, i.e. avoiding the necessity of using magnetic fields. The potential
for practical applications of this effect has been recognized very early 
and includes current driven Magnetic Random Access Memories and Radio Frequency components.

The theoretical description of this physics is usually done in two (relatively independent) steps
where one treats separately the magnetic and transport degrees of freedom.
One first considers the transport degrees of freedom (electrons around the Fermi surface)
and calculates for a given configuration of the magnetization the current and spin current
flowing through the system. In a second step, one considers the magnetization dynamics.
The divergence of the spin current is then identified as a source torque term for the dynamics 
of the magnetic degrees of freedom and is added to the equation of motion of the magnetization. This physical 
picture, sometimes associated with the sd model (the light s electrons are responsible for transport
while the more massive d electrons carry the magnetization) is justified by the difference of time scales
associated with the fast transport degrees of freedom and the slow magnetic degrees of freedom. 
In its simplest form, the magnetization dynamics of a mono domain 
ferromagnetic layer is described by the (phenomenological)  Landau-Lifshitz-Gilbert (LLG) equation
to which one adds the terms calculated from the study of the transport degrees of 
freedom~\cite{slonczewski1996}.

 The first 
corrections to this picture were considered recently~\cite{tserkovnyak2002} and lead to an enhanced magnetic 
damping constant due to the transport electrons, a feature that has been known experimentally 
for quite a long time~\cite{hurdequint1988}: the presence of non magnetic layers sandwiching a magnetic one induces
a large increase of the damping measured in ferromagnetic magnetic resonance spectrum as spin current
flows in and out of the magnetic layer. 
The aim of this paper is to study theoretically the simplest system where both magnetic and transport 
degrees of freedom can be studied on the same footing. Starting from a microscopic model
containing only the original electrons we derive an effective equation for the magnetization
dynamics. The standard concepts of spintronics such as spin accumulation or spin torque 
appear within this formalism and can be viewed as different aspects of the same physics.

The simple system mentioned above is not a mere toy model of academic interest but is actually
realized experimentally in tunneling experiments with ultrasmall magnetic grains. 
Those nanomagnets (typically $3-4 \ nm$ thick grains of magnetic materials
with a total spin $S_0\sim 1000$) are actual model systems for the study of
the interplay between transport and magnetism.  They are small enough so that 
their magnetization can be considered as a large macrospin without spatial
variations~\cite{jamet2001}. Indeed the magnetic exchange length is of the
order of 7 nm in cobalt so that degenerate magnons with non-zero
momentum can be ignored.
Moreover, their mean level spacing is of the
order of $1 meV$,  bigger than the temperature so that
the discreteness of the grain energy spectrum can be
resolved~\cite{gueron1999}. A first set of experiments concentrates on the
transport properties using single electron tunneling spectroscopy (connecting the grain to electrodes
with tunnel junctions) to probe the grain~\cite{gueron1999,deshmukh2001}. The differential
conductance-voltage characteristic of such a device displays peaks
corresponding to excitations of the grain from which the grain's
spectrum can be measured~\cite{kleff2001,kleff2002}.
Another set of experiments focus on the magnetic degrees of freedom of the grain by a direct measure
of the magnetization using a microSQUID technique ~\cite{jamet2001,thirion2003}. Reversal
of magnetic moments as small as 1000 $\mu_B$ have been observed providing
information on the static magnetic
properties of the system (the measure of the switching field as a function of the
external field direction allows the reconstruction of the
anisotropy tensor) and also on the dynamical magnetic
properties (switching times).

The actual device considered in this paper is shown in Fig.~\ref{fig:system} and consists of a nanomagnet 
connected via tunneling barriers to magnetic contacts whose magnetization can be
non-collinear with the grain's. We use this system to illustrate various aspects of
spintronics devices and in particular, current induced magnetic reversal~\cite{slonczewski1996}, current induced spin waves
excitation~\cite{slonczewski1999}, spin accumulation~\cite{valet1993}, Tunneling magneto-resistance~\cite{slonczewski2005,fuchs2004,huai2004} and magnetic damping through the 
conducting electrons~\cite{tserkovnyak2002}. A 3 nm nanograin is not equivalent to a thin
layer however, and its smallness makes its physics more complex. We find that the presence of
the Coulomb blockade phenomena\cite{beenakker1991} strongly affects the spin dynamics.
Its interplay with the magnetic degree of freedom leads to a ``magnetic blockade''
where electrons with a certain spin cannot enter the grain. We show that
the phenomenological LLG equation does not apply to the
nanomagnet and has to be replaced by a new one ( Eq.(\ref{eq:rocks}) which together with its corresponding
phase diagram Fig.~\ref{fig:nanoLLG} is
the main result of this paper) which incorporates the
complex interplay between charge and spin dynamics.
This equation, that we refer as the \nanoLLG equation (Nano Magnet Dynamics), 
is not deduced from phenomenological considerations but rather derived from
a microscopic quantum model. The derivation is interesting in itself as it sheds
light on the origin of the time scale separation between magnetic and transport degrees of freedom.
Early results on the subject were obtained in  Ref.~\cite{waintal2003}, 
and some of the results presented in this paper were announced in a
short publication \cite{PRLWaintalParcollet2005}.

The paper is organized in such a way as to provide two ways of reading. The reader
not interested in technical details can jump directly to the introduction of
 section \ref{sec:Derivation_Nanotorque} where the \nanoLLG equation is presented and then to
section ~\ref{Sec:NanoTorquePhaseDiag} where we study comparatively the phase diagrams of
the LLG and \nanoLLG equation. In section \ref{sec:model}, we
introduce our microscopic model for the nanomagnet. The magnet is described by
a simple Stoner like model and is connected to (non-collinear magnetic) leads via tunneling barriers. An additional
phenomenological coupling to an (bosonic) environment is introduced to correctly describe
the spin relaxation of the grain (via the coupling to phonons, spin-orbit coupling,...).  
In section \ref{sec:QME} we derive the
Master equation for the grain's charge and spin dynamics in close analogy 
to the one found in the standard theory of Coulomb blockade~\cite{beenakker1991}. 
In section, \ref{sec:TMR} we apply this master equation to the Tunnelling Magneto-Resistance (TMR) 
and discuss how it could be used to get extra information on the spectroscopy of the grain. 
While the rest of this paper
is devoted to magnetization dynamics, this section on charge dynamics is relatively self-contained and can be
read independently by readers familiar with tunnelling spectroscopy of ultrasmall grains.
Section \ref{sec:Derivation_Nanotorque} is devoted to the derivation of
the \nanoLLG equation in the semi-classical limit of a large grain's macrospin.
We show that the \nanoLLG equation describes in a unified way the
current induced torque and the mechanism of magnetic relaxation through
the metallic leads.
Section~\ref{Sec:NanoTorquePhaseDiag}, describes in details the phase diagram
of the \nanoLLG equation together with the LLG equation for comparison.

\section{Microscopic model}
\label{sec:model}

\begin{figure}
\vglue +0.45cm
\includegraphics[width=8cm]{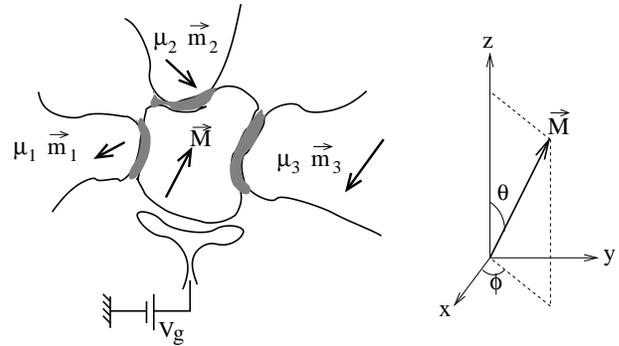}
\caption{\label{fig:system} Left: schematic of the system. Several magnetic reservoirs of
chemical potential $\mu_i$ and magnetization directions $\vec m_i$ (in arbitrary directions)
are connected to the magnetic grain with tunneling barriers. The grain of magnetization
$\vec m$ and easy axis $\vec z$ is also capacitively coupled to a gate at a potential
$V_g$. Right: The magnetization $\vec m$ can be written in spherical coordinates $(\theta,\phi)$}
\end{figure}

Our nanomagnet is weakly connected  through tunneling barriers to
several magnetic electronic reservoirs, whose chemical potential, temperature and magnetization
are respectively $\mu_i$, $T_i$ and $\vec m_i$, and is also capacitively coupled to a gate as shown 
in Fig.\ref{fig:system}. Such a geometry corresponds to the experimental 
geometry of Ref.~\onlinecite{deshmukh2001}.
We neglect any spatial variation of the magnetization in the grain due to 
the large magnetic exchange length (compared to the size of the nanomagnet). In order to describe properly
the relaxation properties of the grain 
(i.e. those not due to electronic transport through the leads), 
we introduce a second coupling term to an environment described by bosonic degrees
of freedom (not displayed in Fig.\ref{fig:system}). This bosonic bath represents all the intrinsic 
degrees of freedom that can provide spin relaxation (phonons, non uniform magnons, impurities...).
As we shall see the presence of this term will lead to a Gilbert type damping term in the magnetization
dynamics.  
Other types of relaxation mechanism that conserve spin have not been included although their presence
might modify our results quantitatively (they affect the charge dynamics). 
The presence of non-equilibrium peaks in the spectroscopy
experiments~\cite{kleff2001,kleff2002} is however a strong indication that charge relaxation rate is 
fairly low in those systems.

\subsection{Isolated magnetic grain}

We start our discussion by introducing a simple Hamiltonian that describe the isolated nanomagnet
with an easy axis along the z axis. 
It is similar to the one discussed in~\cite{canali2000,kleff2001}, and consists of a charge (c) and 
spin (s) part,
\begin{subequations}
  \begin{align}\label{eq:Hdot}
\Hdot &\equiv \Hdot^{c} + \Hdot^{s},
\\
\Hdot^{c} &\equiv \sum_{\alpha\sigma} \epsilon_\alpha
d^\dagger_{\alpha \sigma } d_{\alpha \sigma } + E_C(N-N_{\rm g})^2,
\\
\Hdot^{s} &\equiv - J \vec S\cdot\vec S -\frac{\kappa}{S_0} S_{z}^{2} -\hbar\gamma H S_z.
\end{align}
\end{subequations}
Here, $d^\dagger_{\alpha \sigma }$ ($d_{\alpha \sigma }$) creates (destroys)
an electron on a level $\alpha$ with energy $\epsilon_\alpha$ and
spin $\sigma$ along the z-axis. $N$ is the number of electrons in the grain, $E_C$ the
charging energy and $N_g$ is proportional to the gate voltage. 
The exchange energy ($\propto J$), uniaxial anisotropy ($\propto \kappa$) and Zeeman coupling ($\propto H$,
we consider only a magnetic field along the z axis) are function of the total spin $\vec S$, whose 
expression in term of the electrons present in the system reads :
\be
\vec S = \frac{1}{2} \sum_{\alpha,\sigma_1,\sigma_2}
  d^{\dagger}_{\alpha\sigma_1} \vec\sigma_{\sigma_1\sigma_2}
  d_{\alpha\sigma_2}.
\ee
For $\kappa=0$ and $J$ smaller than the mean level spacing at the Fermi level,
this model is known as the ``Universal Hamiltonian''~\cite{kurland2000,aleiner2002} and has been widely
used to describe quantum dots on energy scales below the Thouless energy. It
corresponds to the limit of zero dimensions (i.e. where the physical
quantities are dimension independent) which is obtained when all the
time scales involved are larger than the time to explore the entire volume
of the grain. When $J$ is
increased, the system undergoes a ferromagnetic (Stoner) instability and acquires a finite
spin $S_0$. The eigenstates $\ket{A}$ of $\Hdot$ of energy $E_{A}$
are readily found by observing
that the charge and spin parts of the Hamiltonian commute with each other so that the
number $n_\alpha=\langle \sum_{\sigma}
d^\dagger_{\alpha \sigma } d_{\alpha \sigma }\rangle=0,1,2$ of electrons in level $\alpha$, the total spin $S$, and the
spin $S_z$ along the easy axis $z$ of the dot are good quantum numbers.

\begin{figure}[htb]
\vglue +0.45cm
\includegraphics[width=6cm]{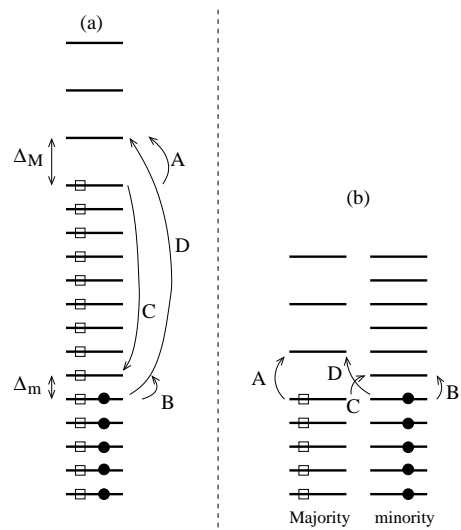}
\caption{\label{fig:excited} (a): Occupation of the different one-body level
in the ground state of the system. Squares (resp. circles) represent the majority
(minority) electrons with mean level spacing of $\Delta_M$ ($\Delta_m$). A,B,C
and D represent low energy excitations, see text. (b): idem but the exchange energy
has been taken into account in addition to the one body energy to show explicitly that
the spin accumulation excitation (C and D) have energies of the same order
as the particle-hole excitations (A and B).}
\end{figure}

 The analysis of the low energy spectrum of $\Hdot$ has been done extensively in~\cite{canali2000,kleff2001}. Here, we reproduce the
main results for the sake of clarity. (i) The charge part of the ground state is
shown in Fig.\ref{fig:excited} (a): $N_s=\sum_\alpha n_\alpha (2-n_\alpha)$ one-body levels are singly occupied in order
for the system to have a non zero spin $S_0=N_s/2$. The Fermi level is then split
into majority and minority electrons, with different mean level spacings,
respectively $\Delta_M$ and $\Delta_m$. (ii) It costs a huge amount of
energy $\propto J S_0$ to switch the spin of one electron without changing the
filling factors ($n_\alpha$) of the one-body levels. $J S_0$ is of the order of the Curie
temperature (around 1000K) which is very large compare to all other energy scales
considered here. Hence we can safely ignore all the excited states with $S<N_s/2$.
We thus assume, for each eigenstate that $S=N_s/2$.
For these low energy excited states, there is no additional degeneracy, and an
eigenstate is entirely described by the set of $n_\alpha$ and $S_z$:
\be
\ket{A}=\ket{ \{n_\alpha\},S_z}
\ee
(iii) there are {\it three} types of low energy excitations, as
sketched in Fig.\ref{fig:excited} (a) and (b).
Following~\cite{kleff2002}, we shall call them particle-hole excitations
for majority (A) and minority (B) electrons,
spin accumulation (C and D: a particle jump from a majority (minority) level to a
minority (majority) level hence changing S by one unit) and spin precession excitation
(E, not shown in the figure: $n_\alpha$ and $S$ are unchanged but $S_z$ is changed).
We note that the huge amount of {\it one-body} energy ($\propto J S_0$) that the
D excitation costs  is almost
exactly canceled by the gain in exchange energy, so that finally all these low energy
excitations are of the same order $J\sim\Delta_M\sim\Delta_m\ll J S_0$. The spin
precession excitation have an energy of order $\kappa$. (iv) the mean level spacing and
exchange energy scale as the inverse of the grains' volume while the anisotropy
parameter $\kappa$ is volume independent. For the grains considered in Refs~\cite{gueron1999,deshmukh2001}
($3 nm$, around $1000$ atoms) however, both have the same order of magnitude
($\kappa\approx 200mK$ and $\Delta_m < \Delta_M$ of a few Kelvin).

We note that although we start with only electronic degrees of freedom, we end up
with low excited states of charge type (A,B,C, and D) and magnetic type (E). The dynamical
interplay between the two will be induced by the relaxation channel (coupling to the bosonic bath)
and tunneling coupling to the leads.
\subsection{Coupling the nanomagnet to the leads and the environment}

Once the isolated grain is understood, we introduce the coupling to the leads
and to the environment. The full Hamiltonian of the systems
reads :
\begin{subequations}
  \begin{align}\label{HamBase1}
H &= \Hdot  + H_{\text{leads}} + H_\text{t} +  H_{\text{env}} +  H_{\text{g}}
\\
H_\text{t} &= \sum_{ak\sigma\alpha} t_{ak\alpha} c^{\dagger}_{ak\sigma}
d_{\alpha \sigma } + \hc
\\
H_{\text{lead }} &= \sum_{ak\beta} \epsilon_{k\beta}
c^{\dagger}_{ak\beta} c_{ak\beta}
\\
 H_{\text{g}} &= \couplingbath \bigl ( S^{+} \phi + S^{-}\phi^{\dagger} \bigr)
\end{align}
\end{subequations}
In this expression, $c^{\dagger}_{ak\sigma}$ ($c_{ak\sigma}$) create (destroys) an electron in
lead $a$ with momentum $k$ and spin $\sigma$ along the z-axis.
The (possibly) ferromagnetic leads are described at the mean field level, using different energies $\epsilon_{k\beta}$
for minority and majority electrons. Here, the index $\beta=M,m$ (Majority,minority) is understood along
the actual magnetization of the lead, which is not necessarily collinear with the easy
axis of the grain (z). The $c^{\dagger}_{ak\sigma}$ are related to the
$c^{\dagger}_{ak\beta}$ (that creates electrons along the magnetization of the leads)
through a $SU (2)$ rotation matrix 
$c^{\dagger}_{ak\sigma}=\sum_\beta R^a_{\sigma\beta} c^{\dagger}_{ak\beta}$ with 
$R^a= \exp[-i\theta_a \vec\sigma\cdot\vec n_a/2]$.
$\theta_a$ is the angle between the magnetization $\vec m_a$ and the z axis,
while $\vec n_a =\vec z\times\vec m_a/|\vec z\times\vec m_a|$.
We assume that no spin-flip takes place during the tunneling processes.

$H_{\text{env}}$ describes the environment in term of a bosonic field
$\phi$. It is an effective, phenomenological description of various
intrinsic relaxation processes of the dot, like  spin-orbit
scattering or phonon-magnon scattering that couple the grain's spin to
bosonic degrees of freedom (like  phonons or non-uniform magnons). A detailed study of 
these processes is beyond the scope of this paper. We restrict ourself to
a phenomenological description of intrinsic damping mechanisms through
spectral density of the boson $\phi$, which will
be denoted by $\rho_{B} (\omega)$. We assume that $\rho_{B}$ is wide
and slowly varying on the frequencies scales at which the  spin
precesses so that the environment do not introduce new time scales in the problem. 
Moreover, we restrict ourselves to a bath coupled to the spin part only, with a coupling
constant $\couplingbath$. We do not introduce a coupling to $S^{z}$ (which do not provide relaxation 
and hence would play no role in the following) or to the charge degree of freedom (we suppose that
the corresponding relaxation time is long as indicated by the presence of non-equilibrium peaks in
the tunneling experiments~\cite{deshmukh2001}). 



\section{Master Equation.}
\label{sec:QME}

The coupling to the leads and environment induces the dynamics of the grain's degrees of freedom.
Here, we use a direct extension~\cite{waintal2003} of the theory of Coulomb blockade in the sequential 
tunneling~\cite{beenakker1991} regime. The dynamics is described by the Master equation for the
probability $P_A(t)$ for the system to be in state $|A\rangle$ at time $t$. This equation has
a simple physical interpretation: the different (tunneling and interacting with the environment)
 events occur incoherently (with rates $\propto\Gamma_a^\sigma$  and $\propto\Gamma_B$ respectively)
 and induce a random walk for the state of the system. The master equation is the equation of conservation 
of probability for this process. It reads, 
\begin{widetext}
  \begin{multline}
    \label{eq:master}
\frac{\partial P_A}{\partial t}=
\sum_{A',a,\alpha\sigma }
\Biggl \{ \Gamma_{a\alpha}^\sigma \left| \langle A' |
  d^{\vphantom{\dagger}}_{\alpha\sigma} | A \rangle \right|^2
\biggl[ n_a\bigl(\Delta E\bigr) P_{A'} - \Bigl( 1 - n_a\bigl(\Delta
E\bigr) \Bigr) P_{A} \biggr] + \Gamma_{a\alpha}^\sigma \left|\langle A'
  | d^{\dagger}_{\alpha\sigma} | A \rangle \right|^2 \biggl[
-n_a(-\Delta E) P_{A}
\\
+ \Bigl(1 - n_a\bigl(-\Delta E\bigr)\Bigr) P_{A'} \biggr] \Biggr\}
+ \pi\couplingbath^2 \!\!\sum_{A',\epsilon = \pm} \!\!\epsilon \left|\langle A'
  | S_x-i\epsilon S_y | A \rangle \right|^2 \biggl[ \rho_B(\epsilon
\Delta E) n_B(\Delta E) P_{A'} + \rho_B(\epsilon\Delta E) n_B(-\Delta
E) P_A \biggr]
  \end{multline}
\end{widetext}
where $\Delta E = E_A - E_{A'}$ is the energy difference between state
$|A\rangle$ and $|A'\rangle$, $n_a(E)=1/(1+e^{(E-\mu_a)/kT})$
($n_B(E)=1/(e^{E/kT}-1)$) is the Fermi (Bose) function, $T$ the
temperature and $\mu_a$ the chemical potential of lead $a$.  
$\Gamma_{a\alpha}^\sigma$ is the tunneling rate (see Eq.~(\ref{eq:gammaeffectif}) and Eq.~(\ref{eq:FGR})
for the precise definition). The term proportional 
to $\Gamma_{a\alpha}^\sigma$ in Eq.(\ref{eq:master}) corresponds to the usual
master equation for tunneling processes, while the second term $\propto \Gamma_B$
is the corresponding term for bosonic degrees of freedom.

As we shall see, the validity of the master equation
approach imposes that the time between two different events
($\hbar/\Gamma_{a\alpha}^\sigma$ and $\hbar/\Gamma_B$) is large compared to the 
coherence time of the leads/environment $\hbar/(kT)$.
The extension to a magnetic grain is not straightforward however as the presence of non-collinear
magnetic contacts could lead to off-diagonal terms in the density matrix
(there is no good quantization axis for the spin). We find that the master
equation remains valid when internal precession is fast enough to
average out contributions transverse to the easy axis. The time needed
for $S_z$ to change significantly is of order $S_0/\Gamma$ (one
tunneling event can only change $S_z$ by half a unit) so that this
condition implies $\Gamma\ll\kappa S_0/\hbar$.

In this section, we perform two tasks. First, we give a proper derivation of
Eq.(\ref{eq:master}) and discuss its domain of validity. Secondly, we evaluate it for
the Hamiltonian of our system Eq.(\ref{HamBase1}) and arrive at the specific form for the
magnetic grain, equations (\ref{FormGalMastereq}), (\ref{eq:master-t}) and
(\ref{master.discrete.bain}).

\subsection{Quantum master equation}
Our starting point for the dynamics of the
system is the quantum master equation for the reduced density matrix $\hat\rho$
of the dot, as derived for instance in \cite{KuboBook} : 
a generic system $a$ (the magnetic grain) coupled to an environment $b$ 
(the leads and the bosonic environment) is described by a Hamiltonian 
$H_{\text{tot}} = H_a + g_{ij} A_i B_j + H_b$ 
where the operators $A$ and $B$ act on  $a$ and $b$ respectively. The quantum 
master equation for the reduced density matrix $\hat\rho$ (where the environment 
degrees of freedom have been traced out) is obtained by 
writing the evolution equation of the full density matrix in second order in the
couplings $g_{ij}$. The trace on the environment can then be taken and one get
an effective equation for the system evolution, 
$$
\partial_t \hat\rho = - \zeta g_{ij} g_{kl}
\int_{-\infty}^t d\tau \ \bigl[A_i(t), A_k(\tau)\hat\rho(\tau)\bigr] \moy{ B_j(t) B_l(\tau)}
$$
\begin{equation}\label{eq:Master_General}
+\bigl[\hat\rho(\tau) A_i(\tau),A_k(t)\bigr] \moy{ B_j(\tau) B_l(t) }
\end{equation}
where $\zeta= 1,-1$ if the operators $A$ and $B$ commute or anticommute. 
If the environment is much larger than the system, we can assume that it is in thermal 
equilibrium and calculate the correlation functions for the $B_j$ operators accordingly
(note that this is not true for the system since it is microscopic and its 
out of equilibrium steady state density matrix is determined by its couplings 
to the environment, even at order zero in those couplings~\cite{parcollet2002}). In the following,
$\moy{\cdots}$ stands for the thermal average of the lead.
Given the form of the microscopic Hamiltonian Eq. (\ref{HamBase1}),
at second order in the couplings, the master equation has the form :
\begin{equation}\label{FormGalMastereq}
\partial_{t} \dst = {\cal D}_{\text{Leads}} \dst  + {\cal
D}_{\text{env}} \dst
\end{equation}
with ${\cal D}_{\text{Leads}} \dst  \sim O (t^{2})$ and ${\cal D}_{\text{env}}
\dst  \sim O (\couplingbath^{2})$ and we can compute the effect of the 
electronic leads and the
bosonic bath separately.

\subsection{Coupling to the leads}
First, let us derive the first term ${\cal D}_{\text{Leads}}
\dst$ of (\ref{FormGalMastereq}).  
Expanding Eq. (\ref{eq:Master_General}) in the tunneling matrix elements, 
we obtain :
\begin{multline}\label{Kubo}
\partial_{t} \dst = -\frac{1}{\hbar}
\sum_{\alpha\alpha'\sigma\sigma'}\int_{-\infty}^{t} d \tau
\biggl \{
\left[ d_{\alpha \sigma} (t), d^{\dagger}_{\alpha' \sigma '} (\tau) \dst(\tau
)\right] A_{\alpha \sigma}^{ \alpha' \sigma' } (t-\tau)\\
+ \left[ \dst(\tau) d^{\dagger}_{\alpha' \sigma '} (\tau)  ,
d_{\alpha \sigma} (t)\right] B_{\alpha \sigma}^{ \alpha' \sigma' } (t-\tau)
+ \hc
\biggr \}
\end{multline}
with the definitions,
\begin{subequations}
  \begin{align}\label{DefAB}
A_{\alpha \sigma}^{\alpha' \sigma' } (t-\tau) &\equiv \frac{1}{\hbar} \sum_{ak}
t_{a k\alpha } t^{*}_{a k\alpha'}
\moy{c^{\dagger}_{a k\sigma } (t-\tau) c_{a k\sigma'} (0)}
\\
B_{\alpha \sigma}^{\alpha' \sigma' } (t-\tau) &\equiv \frac{1}{\hbar} \sum_{ak}
t_{a k\alpha} t^{*}_{a k\alpha'}
\moy{c_{a k\sigma'} (0) c^{\dagger}_{a k\sigma } (t-\tau) }
\\
d_{\alpha \sigma} (t) &\equiv e^{i \Hdot t} d_{\alpha \sigma} e^{- i \Hdot t}
\\
c_{a k\sigma} (t) &\equiv e^{i H_{\text{leads}} t} c_{a k\sigma}
e^{- i H_{\text{leads}} t}
\end{align}
\end{subequations}

Eq.(\ref{Kubo}) in general is rather
complicated since the derivative of $\dst$ at time $t$ involves the density
matrix at former times. Hence the system has ``memory''. The leads correlation
functions $A_{\alpha \sigma}^{\alpha' \sigma' } (t)$
and $B_{\alpha \sigma}^{\alpha' \sigma' } (t)$ can be calculated and behaves typically
like
\be
A_{\alpha \sigma}^{\alpha' \sigma' } (t) \propto \left\{
\begin{array}{l}
\text{constant} \ \ \ \ t<\hbar/\epsilon_F\\
1/t       \ \ \ \  \hbar/\epsilon_F<t<k_BT  \\
e^{-\pi k_BT t/\hbar} \ \ \ \  t\gg k_BT
\end{array}
\right.,
\ee
where $T$ is the temperature of the reservoirs and $\epsilon_{F}$ the
Fermi energy. Therefore this ``memory'' has a correlation time of $\hbar/k_BT$. If the density evolves
on time scales longer than that, $\dst (\tau)$ can safely be replaced by $\dst (t)$
in Eq.(\ref{Kubo}). This is known as the Markovian approximation \cite{KuboBook} and is valid for
$\hbar/kT\ll\hbar/\Gamma$ where $\Gamma$ is the finite width of the grains' levels
due to the coupling to the leads%
(calculated using the Fermi Golden Rule, see the formal
definition in Eq.~\ref{eq:FGR}). 
The domain of validity of the Markovian approximation 
($k_B T \gg \Gamma$) is consistent with the fact that
Eq.(\ref{Kubo}) is obtained by formally expanding in the hopping
amplitude, and hence $\Gamma$ should remain the  smallest energy scale
of the problem. In the other limit, $k_B T \ll \Gamma$, we have
$A_{\alpha \sigma}^{\alpha' \sigma' } (t) \propto 1/t$, the integration
over $\tau$ leads to a divergence and Eq.(\ref{Kubo}) breaks down
entirely.
Once the Markovian approximation has been done, we project
Eq.(\ref{Kubo}) on the many-body eigenstates of the isolated dot, and carry out
the integration over $\tau$. Noting $\dst_{AB}\equiv \bra{A} \dst \ket{B}$ the projection of the density matrix, we obtain

\begin{widetext}
\begin{align}\label{Kubo.proj.z2}
\nonumber
{\cal D}_{\text{Leads}}
\dst_{AB} =
  \sum_{C,D,\alpha\sigma\sigma'}
 &-  e^{i ( E_{A} - E_{D} )t} \dst_{DB}
   \bra{A}d_{\alpha\sigma}\ket{C} \bra{C}d^{\dagger}_{\alpha \sigma'}\ket{D}
   A_{\alpha \sigma}^{\alpha \sigma'} ( E_{C} - E_{D})
\\
\nonumber
 &-  e^{i ( E_{C} - E_{B} )t} \dst_{AC}
   \bra{C}d^{\dagger}_{\alpha\sigma'}\ket{D}
\bra{D}d_{\alpha\sigma}\ket{B}
   B_{\alpha \sigma}^{\alpha \sigma'} (E_{C} - E_{D})
\\
\nonumber
 &+ e^{i ( E_{A} - E_{C} + E_{D} - E_{B} )t}  \dst_{CD}
   \bra{A}d^{\dagger}_{\alpha\sigma'}\ket{C}  \bra{D}d_{\alpha\sigma}\ket{B}
   A_{\alpha \sigma}^{\alpha \sigma'} ( E_{A} - E_{C})
\\
 &+ e^{i ( E_{A} - E_{C} + E_{D}
 - E_{B} )t}  \dst_{CD}
   \bra{A}d_{\alpha\sigma}\ket{C} \bra{D} d^{\dagger}_{\alpha\sigma'}\ket{B}
   B_{\alpha \sigma}^{\alpha \sigma'} ( E_{D} - E_{B})
+ (A \leftrightarrow B)^*
\end{align}
\end{widetext}
where we introduced
\begin{equation}
A (E) \equiv \int_{0}^{\infty } dt \,\, e^{-iE t}A (t)
\end{equation}
Eq.(\ref{Kubo.proj.z2}) reduces to the (first part of the) master equation 
Eq.(\ref{eq:master})  if we ignore completely the off diagonal part
of the density matrix and note  $P_{A}=\delta_{AB}\dst_{AB}$ its diagonal part. Such an approximation,
although very common in Coulomb blockade theory needs to be justified in the present context.
Indeed, when the magnetizations of the reservoirs are non-collinear with
the easy axis of the grain, neglecting the off diagonal part of the density matrix amongst to 
neglect all the dynamics in the $xy$ plane (as the matrix elements $\dst_{AA}$ depend only on the $z$
component of the magnetization).
 In other words, in this approximation an electron tunneling along the x axis
is equivalent to an electron tunneling along the y axis: if two reservoirs have their 
magnetization lying in the $xy$ plane, then the physics is independent of the relative angle 
between their two magnetizations. A case where the off-diagonal elements of the density
matrix have to be considered has been discussed in~\cite{braig2005}.

On a qualitative level, we find that the off-diagonal part of the density matrix
oscillates at a frequency set by the excitation energies (the phases in Eq.(\ref{Kubo.proj.z2}))
while the density matrix evolve on a rate $\Gamma$. Hence, when the excitation energies
are larger than $\Gamma$, the off-diagonal density matrix averages to zero.
For our model, this translates into $\kappa\gg\Gamma$. In this limit,
which is achieved for instance in the grains studied in ~\cite{gueron1999,deshmukh2001}, 
the magnetization
of the grain precesses several times around its easy axis ($z$) in between two tunneling
events, so that the transverse degrees of freedom (along $xy$) are averaged out, and the only
relevant quantity is the projection of the magnetization along the $z$ axis.
In this limit,
all the physics of non-collinear reservoirs
can be understood {\it as if the magnetizations of the reservoirs were all
lying on the $z$-axis}.
This argument can be made more precise as explained in Appendix A. The master equation
is of course also valid when the leads magnetization are parallel to the easy axis z.

The above argument is rather general, and does not depend on the particular form of
$\Hdot$. As we shall see below, the magnetic grain displays a separation between
the charge and spin relaxation times (respectively of order $1/\Gamma$ and
$S_0/\Gamma$). This shall increase the range of validity of our approach to
$\kappa\gg \Gamma/S_0$ (see below section \ref{sec:Derivation_Nanotorque}).

Eq.(\ref{Kubo.proj.z2}) reduces to the first part of Eq.(\ref{eq:master}),
\begin{widetext}
\begin{align}
\label{eq:master2}
{\cal D}_{\text{Leads}}
\dst_{A} =
\nonumber
 \sum_{C,a\alpha\sigma}   \Gamma_{a\alpha}^{ \sigma}
\biggbra{
 &|\bra{A}d_{\alpha\sigma}\ket{C}|^2
 \Bigparent{
 \dst_{C} \check n_a( E_{C} - E_{A}) - \dst_{A}  n_a( E_{C} - E_{A})
 }
\\
+
&|\bra{A}d^{\dagger}_{\alpha\sigma}\ket{C}|^2
 \Bigparent{
   \dst_{C} n_a( E_{A} - E_{C})
-
   \dst_{A} \check n_a(E_{A} - E_{C})
}
}
\end{align}
\end{widetext}
where $\check n_{a} (x)\equiv 1-n_{a} (x)$ and $\Gamma_{a\alpha}^{ \sigma}$ is the effective tunneling rate from reservoir $a$ on
the state $\alpha$ along the z-axis. It is connected to the tunneling rate along
the magnetization direction $\vec m_a$ of reservoir $a$ through,
\begin{multline}
\label{eq:gammaeffectif}
\Gamma_{a\alpha}^{ \uparrow}=\cos^2 \frac{\theta_a}{2}\Gamma_{a\alpha}^{U}
+\sin^2 \frac{\theta_a}{2}\Gamma_{a\alpha}^{D}
\\  
\Gamma_{a\alpha}^{ \downarrow}=\cos^2 \frac{\theta_a}{2}\Gamma_{a\alpha}^{D}
+\sin^2 \frac{\theta_a}{2}\Gamma_{a\alpha}^{U}
\end{multline}
where $\theta_a$ is the angle between the $z$ axis and $\vec m_a$, while
$\Gamma_{a\alpha}^{ \beta}$ ($\beta=U,D$) is given by the Fermi Golden Rule:
\be
\label{eq:FGR}
\Gamma_{a\alpha}^{\beta}(\epsilon)\equiv \frac{2\pi}{\hbar}
\sum_k  |t_{ak\alpha}|^2 \delta(\epsilon-\epsilon_{ak\beta})
\ee
 Eq.(\ref{eq:master2}) was used previously for reservoirs whose magnetization was parallel
to the easy axis of the grain. Here its regime of validity is extended
to non-collinear reservoirs. 

The average electrical current flowing from reservoir $a$ onto the grain reads,
$$
I_a =
 e \sum_{A,C,\alpha\sigma} \dst_{C}  \Gamma_{a\alpha}^{ \sigma}
\biggbra{
 |\bra{A}d^{\dagger}_{\alpha\sigma}\ket{C}|^2 n_a( E_{A} - E_{C})
$$
\begin{equation}
  \label{eq:current1}
-|\bra{A}d_{\alpha\sigma}\ket{C}|^2    \check n_a( E_{C} - E_{A})
}  
\end{equation}

We now turn to evaluate Eq.(\ref{eq:master2}) specifically for the case considered
in this paper. We restrict ourself to the Coulomb blockade regime where $E_C$
is large enough so that only states with $N$ and $N+1$ particles need
to be considered, and we note $V_g$ the corresponding charging energy
(which is tunable with the gate voltage). Let us define $\bar\rho_{+} (\{n'_\beta\}, S'_z )$
(resp. $\bar\rho_{-} (\{n_\beta\} ,S_z$)
the density matrix in the sector with $N+1$ (resp. $N$) particles with
$S'=\frac{1}{2}\sum_\beta n'_\beta (2-n'_\beta)$ (resp.
$S=\frac{1}{2}\sum_\beta n_\beta (2-n_\beta)$). The matrix elements,
$\bra{S_z, \{n_\beta\}}d^{\dagger}_{\alpha\sigma}\ket{S'_z, \{n'_\beta\}}$
factorize into a charge part and a
spin part. The charge part, noted $F_\alpha (\{n_\beta\}, \{n'_\beta\})$
is equal to zero except when $n_\beta=n'_\beta$
for every $\beta\ne \alpha$ and  $n'_\alpha=n_\alpha-1$ where its value is one.
The spin part is a Clebsch-Gordon coefficient and reads (Cf. Appendix
C, formulas C.25 and C.26 of \cite{Messiah}),
$$
\left|\moy{ S', S'_z
\Bigl| d_{\alpha\sigma}^{\dagger} \Bigr|  S ,S_z }\right|^{2}
= 
$$
\be\frac{ S + \eta_\alpha \sigma  S_z + \frac{1+\eta_\alpha}{2}}{2S+1}
\delta_{S',S+\frac{\eta_\alpha}{2}}
\delta_{S'_z,S_z+\frac{\sigma}{2}}
\ee
where $\eta_\alpha=+1$ ($\eta_\alpha=-1$) for a majority (minority) state.
After calculating the energy differences (the one body energies are counted from
their respective minority/majority Fermi level),
\begin{subequations}
\begin{align}
\Delta E &= \epsilon_\alpha +V_g -J\frac{1+2\eta_\alpha}{2}
-\frac{\kappa}{S_0} S_z \sigma  - \frac{\kappa}{4S_0} - h \frac{ \sigma}{2}
\\
\Delta E' &= \epsilon_\alpha +V_g -J\frac{1+2\eta_\alpha}{2}
-\frac{\kappa}{S_0} S'_z \sigma  + \frac{\kappa}{4S_0} - h \frac{
\sigma}{2}
\end{align}  
\end{subequations}
we finally obtain
\begin{widetext}\begin{subequations}
  \label{eq:master-t}
\begin{align}
\nonumber
{\cal D}_{\text{Leads}}
\bar \rho_{+} (\{n'_\beta\},S'_z) =
\sum_{a\alpha\sigma\{n_\beta\} } & \Gamma^{\sigma}_{a\alpha} F_\alpha (\{n_\beta\}, \{n'_\beta\})
\frac{ S' + \eta_\alpha \sigma  S'_z + \frac{1-\eta_\alpha}{2}}{2S'+1-\eta_\alpha}
\times \\
&\biggbra{
n_{a}(\Delta E') \bar \rho_{-} \left(\{n_\beta\},S'_z-\frac{\sigma}{2} \right)
- \check n_{a} (\Delta E') \bar \rho_{+} \left(\{n'_\beta\},S'_z \right)}
\\
\nonumber
{\cal D}_{\text{Leads}}
\bar \rho_{-} (\{n_\beta\}, S_z)  =
\sum_{a\alpha\sigma\{n'_\beta\} } &\Gamma^{\sigma}_{a\alpha} F_\alpha (\{n_\beta\}, \{n'_\beta\})
\frac{ S + \eta_\alpha \sigma  S_z +\frac{1+\eta_\alpha}{2}}{2S+1}
\times \\
&\biggbra{
\check n_{a} (\Delta E) \bar \rho_{+} \left(\{n'_\beta\}, S_z + \frac{\sigma}{2} \right)
- n_{a}(\Delta E) \bar \rho_{-} \left(\{n_\beta\},S_z \right)}
\end{align}
\end{subequations}\end{widetext}
It is important to realize that {\it four} types of tunneling
events can occur: The electrons can tunnel  either on a majority or minority state,
with a spin either up or down (with respect to the easy axis z). 
This apparently paradoxal statement just reflects the
fact that we measure the spin along the easy axis and not along the magnetization of the grain.
When the system is near equilibrium, the Clebsch-Gordan
coefficients almost vanish when a up (down) electron tunnel on a minority (majority)
state, making the distinction between minority/majority and up/down electrons meaningless.
However, when the system is put out of equilibrium (as it will be in the presence of a
bias voltage) such a distinction is relevant.

\subsection{Coupling to the bosonic environment}

In this paragraph, we derive the second term ${\cal D}_{\text{env}}
\rho$ of (\ref{FormGalMastereq}).  The quantum master equation for the environment reads using Eq.(\ref{eq:Master_General}) and noting $S^\pm=S_x\pm S_y$:
\begin{align}\label{Kubo.bath}
\nonumber
\partial_{t} \dst =  -& \couplingbath^{2}  \int_{-\infty}^{t} d \tau \Big[S^{+}
(t),S^{-} (\tau) \dst (\tau) \Big] \moy{\phi (0)\phi^{\dagger} (\tau
-t)}
 \\
-& \couplingbath^{2}  \int_{-\infty}^{t} d \tau \Big[S^{-}
(t),S^{+} (\tau) \dst (\tau) \Big]
\moy{\phi^{\dagger}(0)\phi^{}(\tau-t)} + \hc
\end{align}

As in the previous calculation, we  use the Markov approximation, neglect 
the off diagonal terms of the density matrix projected on the eigenstates of the grain.
We obtain the second part of Eq.\ref{eq:master}. The specific form for our problem, reads
\begin{widetext}
\begin{align}\label{master.discrete.bain}
\nonumber
{\cal D}_{\text{env}}
\bar \rho_{\pm} (\{n_{\alpha } \},S_z) = \couplingbath^2
\Biggl (
&  \left|\bra{S_z} S^{+} \ket{S_z-1} \right|^{2}
\biggl(
A_{\phi } (E_{S_z-1} - E_{S_z}) \bar \rho_{\pm} (\{n_{\alpha } \},S_z-1) - B_{\phi } (E_{S_z-1}- E_{S_z}) \bar \rho_{\pm}(\{n_{\alpha } \},S_z)
\biggr)
 \\
+&
 \left|\bra{S_z+1} S^{+} \ket{S_z} \right|^{2}
\biggl(
B_{\phi } (E_{S_z} - E_{S_z+1}) \bar \rho_{\pm} (\{n_{\alpha } \},S_z+1) - A_{\phi } (E_{S_z}- E_{S_z+1}) \bar \rho_{\pm} (\{n_{\alpha } \},S_z)
\biggr)
\Biggr )
\end{align}
\end{widetext}
where we used the definitions
\begin{subequations}
\begin{align}\label{def.bathAB}
E_{S_z} &= - h S_z  - {\frac{\kappa}{S_{0}} S_z^{2}}
\\
A_{\phi } (E) &\equiv  \int_{0}^{\infty} \moy{\phi^{\dagger} (u) \phi (0)}
e^{i E u}\\
B_{\phi } (E) &\equiv  \int_{0}^{\infty} \moy{\phi(u) \phi^{\dagger}  (0)}
e^{-i E u}
\end{align}
\end{subequations}
The imaginary part of $A_{\phi }$ and $B_{\phi }$ only renormalizes the precession
frequencies of the spin \cite{Takahashi_Shibata}, so we will drop
them. After standard manipulations, we reduce (\ref{def.bathAB})
to
\begin{subequations}
\begin{align}\label{bath.AB.reduit}
\Re A_{\phi } (E ) &= \pi  n_{B} (-E )\rho_{B} (-E) \\
\Re B_{\phi } (E ) &= - \pi  \check n_{B} (-E )\rho_{B} (-E)
\end{align}
\end{subequations}
where $n_{B}(E)=1/(e^{E/kT} -1)$ is the Bose function and $\check n_{B} (E) \equiv n_{B} (-E)$.
Note that as the bosonic environment couples only to the spin degree of freedom, 
it is diagonal in $\bar\rho_{+}, \bar\rho_{-}$. 
Equations (\ref{FormGalMastereq},\ref{eq:master-t},\ref{master.discrete.bain}) 
define the master equation that will be used in the rest of this paper.

\section{Grain's spectroscopy and tunnelling magneto resistance}
\label{sec:TMR}
 
One important aspect of the Master equation (\ref{FormGalMastereq},\ref{eq:master-t},\ref{master.discrete.bain})  is its validity
for non-collinear systems. Indeed in real samples, it is very difficult to have
the easy axis of the grain aligned with the magnetization of the contact. 
However, Eq.(\ref{eq:gammaeffectif}) can be interpreted 
{\it as if } 
the anisotropy axis of the grain were aligned with the magnetization direction of the contact.
Hence, the tunnelling magneto-resistance can be measured (in a sample with one magnetic lead)
by switching the direction of the lead magnetization while keeping the magnetization of the
grain unchanged (this could be done for instance using permalloy, whose coercitivity of the
order of 10 mT is much smaller than the anisotropy of the grains, typically 100mT) irrespectively
of the angle between the two magnetizations. As we shall see, those TMR experiments can provide
extra information on the spectroscopy of the grain and in particular should help to differentiate
non-equilibrium peaks associated to spin accumulation excitation from those associated to particle-hole
excitations.

    The current-voltage characteristics of a magnetic grain with non magnetic lead has
been analyzed extensively both theoretically~\cite{kleff2001,kleff2002} 
and experimentally~\cite{gueron1999,deshmukh2001}.
The differential conductance $\partial I/\partial V_{\rm bias}$ shows peaks that allow one
to extract information about the low energy spectrum of the grain 
(assuming without loss of generality that upon applying a bias voltage $V_{\rm bias}$
across the sample the chemical potential of lead $2$ is unchanged while the one of lead $1$
is increased by $e V_{\rm bias}$). The basic rules to get a peak at a $V_{\rm bias}^*$ are
(i) there must be two eigenstates of the grain $|A_N\rangle$ and $|A_{N+1}\rangle$
with respectively $N$ and $N+1$ particles,
such that $E_{A_{N+1}}-E_{A_{N}}=e V_{\rm bias}^*$ (being in $|A_N\rangle$
the system must be able to get enough energy from the lead to reach $|A_{N+1}\rangle$
in a tunneling event). (ii) There must be a non zero matrix element
$\langle A_{N+1} | d^{\dagger}_{\alpha\sigma} |A_{N}\rangle$. (iii) The probability of being
in state $|A_N\rangle$ must be non zero. The case where $|A_N\rangle$ is the ground
state of the grain with $N$ electrons is the simplest to analyze and has been used
to extract the one-body energies $\epsilon_\alpha$ of non magnetic grains~\cite{black1996}.
All the other peaks are referred to as non-equilibrium excitations~\cite{agam1997,oreg2002}. In non magnetic
grains, particle-hole excitations are at the origin of extra peaks, provided the
charging energy has mesoscopic fluctuations, i.e. depends slightly on $|A_N\rangle$.
In addition it was found that in magnetic grains, spin accumulation excitations
also give peaks while spin precession excitations do not.

In order to illustrate the interest of the TMR experiments, we have calculated numerically the
exact solution of Eq.(\ref{FormGalMastereq},\ref{eq:master-t}) without bath and plot in Fig.\ref{fig:MR}
the differential conductance (left panel) and the difference between the current in the aligned
and anti-aligned configuration (right panel) as a function of the bias voltage $V_{\rm bias}$
and gate voltages $V_{\rm g}$ for an illustrative sample. The stationary solution of 
Eq.(\ref{FormGalMastereq}) amounts to solve a linear 
equation of the form $A X =0$ which is efficiently done numerically by separating the diagonal $D$ 
and off-diagonal part $E$ of $A=D-E$ and multiplying iteratively a initial vector $X$ by the matrix 
$D^{-1}E$ until convergence (which is guaranteed by the fact that $A$ is a stochastic matrix\cite{doob1953}). In the left panel for $V_{\rm g}\le \alpha\approx 2$ we see a typical 
spectra where each peak (a,c,and e) is associated to a different (in this case minority) level
(cartoon 1).
For $V_{\rm g}\ge \alpha$, (cartoon 2 and 3) we see that two extra ``non-equilibrium'' peaks appear (b and d). These
peaks are associated to spin-accumulation excitations and have been described in detail in \cite{kleff2002}
(an electron enters on the minority level (a) leaves from a majority level which decreases the exchange
energy of the grain hence the bias voltage needed for an electron to tunnel on the (c) minority level).
On the right panel, we see a similar structure (except that total currents have been plotted not conductances)
with an additional structure at $ V_{\rm g}\le \beta\approx 3$ which is barely visible on the left panel.
This additional structure comes from the fact that there can be more than one spin-accumulation excitation
that gives a peak approximately at the
same bias voltage (in our example the electrons can leave from two different majority levels, cartoon 4) and
while the positions of the two peaks might be hard to resolve, their contribution to the magneto-resistance can be
very different. 
We also note that the sign of the TMR shown in Fig.\ref{fig:MR} is negative almost everywhere (the tunneling happens
on minority levels) except for the upper right corner $V_{\rm g}\ge \beta$ and $V_{\rm bias}\ge e$
where it becomes positive. Hence the sign of the TMR can be controlled, at least for the choices
of parameters used in this example, by external voltages.
\begin{figure}[htb]
\vglue +0.45cm
\includegraphics[width=8cm]{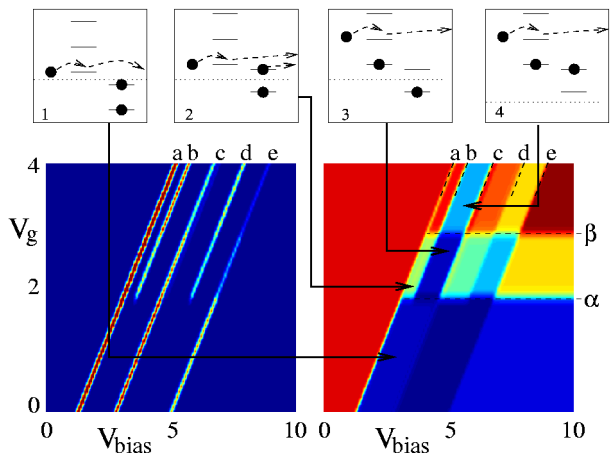}
\caption{\label{fig:MR}    (Color Online).
Differential conductance $dI/dV_{\rm bias}$ (left panel) and difference between the current in the aligned
and anti-aligned configuration $I(\theta=0)-I(\theta=\pi)$ (right panel) as a function of the bias voltage $V_{\rm bias}$
and gate voltages $V_{\rm g}$ for an illustrative sample. The magnitudes (in arbitrary unit) increase when
going from blue to red. The cartoons represent the different type of tunneling events occurring for different
values of $V_{\rm g}$ and $V_{\rm bias}$. The levels on the left of each cartoon represent the minority levels 
while the majority are on the right. The dotted line stands for the Fermi level of the right (non magnetic) contact 
while the electrons entering from the left (magnetic) contact have an energy which is $V_{\rm bias}$ higher.}
\end{figure}


\section{Macroscopic Spin limit} 
\label{sec:Derivation_Nanotorque}
\def\orderUN{O \parent{\Gamma/\Sm}}
\def\orderDEUX{O \parent{\Gamma/\Sm^2}}
\def\orderTROIS{O \parent{\Gamma/\Sm^3}}
In actual nanomagnets, the total spin is large (around $1000$) which considerably
simplifies the master equation into a differential equation for the magnetization dynamics
analogous to the LLG equation.
In this section, we perform this derivation. First, we take the large spin $S_0$ limit
in the master equation, and replace the discrete degree of freedom $S_z$ (projection of the spin
along the easy axis $z$) by its continuum
version $u\sim S_z/S_0$. Second, we find that charge and spin equilibrium occurs on
two different time rates, (fast $\Gamma$ for the charge and slow $\Gamma/S_0$ for the spin) 
so that we can integrate the charge dynamics. The result has the form of a Fokker-Planck
equation for $u$. The associated Langevin equation (the \nanoLLG equation) is the chief result 
of this paper and reads,
\begin{subequations}
\label{eq:rocks}
\begin{align}
\dot{u} &= R (u) + \sqrt{\frac{2 D(u)}{S_0 }} \xi_{t}
\\
R (u) &=  (1-u^{2}) \biggl[
 \alpha_{B} (h + 2 \kappa u)\nonumber  \\  &+
\frac{(1-q) f_{\uparrow} (u) -
(1+q)f_{\downarrow} (u)}{4 S_0  ( 1 + q \eta u)}
\biggr]
\\ \nonumber
D (u) &= (1-u^{2})
\biggl[
\alpha_{B} \frac{h+2\kappa u}{2 {\rm th }\left(\frac{h+2\kappa u}{2kT}\right)}
\\  \nonumber &+
\frac{(1-qu^{2}) f_{\uparrow} (u) + (1+qu^{2})f_{\downarrow} (u)
}{8 S_0 (1 + q \eta  u)}
\\ 
&+\frac{
u^{2}
\bigl (
f_{\uparrow} (u)  - f_{\downarrow} (u)
 \bigr)^{2}
- \bigl (
f_{\uparrow} (u)  + f_{\downarrow} (u)
 \bigr)^{2}
}{8 S_0\Gamma (1 + q \eta  u )}
\biggr]
\end{align}
\end{subequations}
where we have introduced the following notations: the magnetic field along $z$ is 
$h=\gamma H$, $\xi_{t}$ is a white noise 
$\langle \xi_t\xi_{t'}\rangle=\delta(t-t')$
and the Gilbert damping term $\alpha_{B}$ is the intrinsic damping of the isolated grain.
$\eta=\pm 1$ depending on whither the tunneling occurs on a majority/minority state. 
The functions $f_{\sigma}(u)$ (with $\sigma=\pm 1$ for up/down spin)
contain all the dependence in temperature $T$ and leads chemical potentials $\mu_a$,
\be
f_{\sigma}(u)= \sum_a \Gamma_a^{\sigma} n (\epsilon -\kappa\sigma u -h \frac{\sigma}{2}-\mu_a)
\ee
where the offset energy $\epsilon$ includes the charging energy for an electron to enter the grain,
$\kappa$ is the uniaxial anisotropy and $n(E)$ is the Fermi function. For simplicity, we have restricted ourselves to the 
case where the tunnelling events occur on only one one-body level with tunneling rate
$\Gamma_a^\sigma$.
$q$ is the spin-torque asymmetry (or the average polarization of the tunneling rates) and $\Gamma$ the total tunneling rate,
\begin{equation}\label{DefpGamma}
q\equiv\frac{\sum_a \Gamma^\uparrow_a -\Gamma^\downarrow_a}{\sum_a\Gamma^\uparrow_a +\Gamma^\downarrow_a}
\ \ , \ \ \Gamma=\sum_a\Gamma^\uparrow_a +\Gamma^\downarrow_a
\end{equation}

At the classical level, the $z$ component of the spin has a non-linear
deterministic evolution, corrected by a small stochastic noise. The
deterministic part takes a form similar to the LLG equation (See below
Eq.(\ref{eq:llg-z})), except from the current induced term which is
modified by the Coulomb blockade phenomena. The analysis of Eq.(\ref{eq:rocks}) 
will be done in section \ref{Sec:NanoTorquePhaseDiag} while the rest
of this section is devoted its derivation. Hereafter, Equation (\ref{eq:rocks})
will be referred to as \nanoLLG  (Nano Magnet Dynamics).

\subsection{Large spin limit}

As we restrict ourselves to tunneling events on one orbital $\alpha$,
 we drop the index $\alpha$. The system can be in one of two spin states with spin
$S_0$ ($N$ particles) or $S'_0=S_0+\eta/2$ ($N+1$ particles).
For later convenience we perform the large spin expansion in $1/\Sm$ with ${\Sm}
\equiv S_0 + \eta /4$, and denote $\tilde{\kappa}/\Sm \equiv\kappa /S_{0}$. Moreover,
we assume $\couplingbath^{2} =O(\Gamma/\Sm)$ so that the intrinsic spin relaxation rate will be
comparable to the relaxation induced by the presence of the contacts.
In the classical limit, the magnetizations are continuous variables $-1\leq u \leq 1$, such that
$S_{z}' = S'_0 u $ and $S_{z} = S_0 u$. We denote by $\rho_{\pm}$ the
functions (of $u$) such that (normalization factors are chosen for convenience):
\be
\bar  \rho_{+} (S_{z}') = \frac{S'_0 }{\Sm} \rho_{+}
\parent{\frac{S_{z}'}{S'_0} }
\ \ , \ \
\bar  \rho_{-} (S_{z}) =  \frac{S_0}{\Sm} \rho_{-}
\parent{\frac{S_{z}}{S_0} }
\ee
Introducing the notation $x_{\sigma} = \epsilon + V_{g}
-J\frac{1 + 2 \eta}{2} -\tilde{\kappa} u\sigma  - h \sigma /2$,
the master equation (\ref{FormGalMastereq},\ref{eq:master-t},\ref{master.discrete.bain}) 
now takes the form,
\begin{widetext}
\begin{subequations}
\begin{align}\label{MasterEq.Continue}
\nonumber
\partial_{t} \rho_{+} (u)  =& \sum_{a,\sigma }
\Gamma^{\sigma}_{a}
\frac{S'_0\parent{1 + \eta \sigma u} + \frac{1-\eta}{2}}{2S_0+1}
\left[
n_{a}\Bigparent{ x_{\sigma} + \frac{\tilde{\kappa}}{4\Sm} ( 1- \eta \sigma  u) }
\frac{S'_0}{S_0}\rho_{-} \parent{\frac{S'_0 u - \frac{\sigma}{2}}{S_0}}
-
\check n_{a}\Bigparent{  x_{\sigma} + \frac{\tilde{\kappa}}{4\Sm} ( 1-\eta \sigma  u) }
\rho_{+} (u)
\right]
\\
& + {\cal D}_{\text{env}} \rho_{+} (u)
\\
\nonumber
\partial_{t} \rho_{-} (u)  =& \sum_{a,\sigma }
\Gamma^{\sigma}_{a }
\frac{S_0 (1+ \eta \sigma  u) + \frac{1+\eta}{2}}{2S_0+1}
\biggl[
\check n_{a}\Bigparent{  x_{\sigma} - \frac{\tilde{\kappa}}{4\Sm} ( 1-
\eta \sigma  u) }
\frac{S_0}{S'_0}\rho_{+}  \parent{\frac{ S_0 u + \frac{\sigma}{2}}{S'_0} }
-n_{a}\Bigparent{  x_{\sigma} - \frac{\tilde{\kappa}}{4\Sm} ( 1- \eta \sigma  u) }
\rho_{-} (u)
\biggr]
\\
& + {\cal D}_{\text{env}} \rho_{-} (u)
\end{align}
\end{subequations}
We now proceed with the expansion in $1/\Sm$ which assumes that
that the functions $\rho$ varies slowly with $u$, such that
$|\partial_{u} \rho_\pm (u)|\ll |\rho_{\pm} (u)| \Sm$.
We get,

\begin{align}
\label{EqCharge}
\partial_{t} \rho_{+}^{} (u)  =& \sum_{a,\sigma }
\frac{\Gamma^{\sigma}_{a }}{2}
\parent{1 + \eta \sigma u}
\Bigparent{
    n_{a} (x_{\sigma}) \rho_{-}^{} (u)
 - \check n_{a}(x_{\sigma})\rho_{+}^{} (u)
}
+ \orderUN
\\
\label{Eqrhospin}
\partial_{t} \rho (u)  =&  -
\partial_{u} \bigl (
j_{\text{leads}} + j_{\text{env}} \bigr ) +\orderTROIS
\end{align}
\begin{align}
j_{\text{env}} =&
- \pi \couplingbath^{2}  \rho_{B} (-h - 2\kappa u)
\Biggl [
\Bigl(
1-u^{2} + {\frac{1}{\Sm}}
\Bigr)
\rho (u)
 + \frac{(1-u^{2})}{2\Sm }
   \bigl (
 1 + 2 n_{B} (-h - 2\kappa u)
   \bigr )
   \partial_{u} \rho (u)\Biggl ]
\\
\nonumber
j_{\text{leads}} \equiv& \frac{1}{4 \Sm} \sum_{a,\sigma } \Gamma^{\sigma}_{a}
\Biggl[
\Bigparent{ \sigma (1-u^{2}) + \frac{\sigma (1-\eta \sigma u)^{2}}{4\Sm}}
\Bigparent{
    n_{a} (x_{\sigma}) \rho_{-} (u)
 -
\check n_{a} (x_{\sigma}) \rho_{+} (u)
}
\\
\nonumber
& +
\frac{1}{4\Sm}
 \sigma (1-u^2)  
\Bigparent{
   \check n_{a} (x_{\sigma}) \rho_{+} (u)
  + (2\eta -1) n_{a} (x_{\sigma}) \rho_{-} (u)
}
\\
-&
\frac{1}{4\Sm}
(1- \eta \sigma u)(1-u^{2})
\Bigparent{
   \check n_{a} (x_{\sigma}) \partial_{u }\rho_{+} (u)
  + n_{a} (x_{\sigma})\partial_{u} \rho_{-} (u)
}
\Biggr ]
\end{align}
\end{widetext}
where we have introduced $\rho (u) \equiv \rho_{+} (u) + \rho_{-}
(u)$, the reduced density matrix for the spin only, traced over the charge
degrees of freedom.

\subsection{Spin dynamics}

We note that the terms of order $\Gamma$ only couple to the charge degrees
of freedom. The terms that affects the spin ($u$) are of order $\Gamma/\Sm$
so that in leading order,
(and we will be interested in the sub-dominant
order only close to $u=\pm 1$, see below)
the charge degree of freedom is at equilibrium and we have,
(up to terms of order $\Gamma/\Sm^3$)
\begin{subequations}
  \begin{align}
\rho_{+} (u) &= \frac{2}{\Gamma }\frac{a (u) + \eta u b (u)}{1  +  q
\eta u} \rho (u)
\\
\rho_{-}(u) &= \frac{2}{\Gamma }\frac{\check a (u) + \eta u \check b (u)}{1 +
q \eta u} \rho (u)
\end{align}
\end{subequations}
where 
\begin{subequations}
  \begin{align}\label{Def_ab}
a (u) & \equiv  \sum_{a,\sigma } \frac{\Gamma^{\sigma}_a}{2} n_{a} ( x_\sigma)&
\check a (u)& \equiv  \sum_{a,\sigma }
\frac{\Gamma^{\sigma}_{a}}{2} \check n_{a} (x_\sigma)&
\\
b (u)  & \equiv  \sum_{a,\sigma } \sigma
\frac{\Gamma^{\sigma}_a}{2} n_{a} (x_\sigma )&
\check b (u) & \equiv  \sum_{a,\sigma } \sigma
\frac{\Gamma^{\sigma}_a}{2} \check n_{a} (x_\sigma )&
\end{align}
\end{subequations}
(we recall that $\check n(x)\equiv n(-x)$).
Using the charge equilibrium relations, Eq.(\ref{Eqrhospin}) finally takes the form of
a Fokker-Planck equation and reads, 
\def\uu{(u)}
\begin{subequations}
  \begin{align}\label{Fokker-Planck}
\partial_{t} \rho (u)  &=
 \partial_{u}
\biggbra{ - R (u)\rho (u)
+ \frac{1}{\Sm} \partial_{u}\biggparent{D (u) \rho (u)} 
}
\\
R (u) &\equiv  \frac{1}{\Gamma\Sm }(1-u^{2}) \frac{\check a \uu b \uu - a \uu \check b \uu}{1 +
q\eta  u}
\nonumber\\
&- \pi \couplingbath^{2} (1-u^{2}) \rho_{B} (-h - 2\kappa u) +\orderTROIS
\\ \nonumber
D (u) &\equiv  \frac{1}{2\Gamma \Sm  }
(1-u^{2})
\frac{a\uu\check a \uu - u^{2} b \uu
\check b \uu}{1  + q\eta u}
 \\
\nonumber &+
 \frac{\pi  \couplingbath^{2}}{2 }
(1-u^{2}) \rho_{B} (-h - 2\kappa u)
  \times\nonumber\\
&\Bigl ( 1 + 2 n_{B} (-h - 2\kappa u) \Bigr)+ \orderDEUX
\end{align}
\end{subequations}
where $R$ and $D$ have be calculated only at dominant order. The sub dominant correction to $R$
will be studied below.
Equation (\ref{Fokker-Planck}) is in the Ito form~\cite{GardinerBook}, and the
corresponding stochastic differential equation can therefore be identified straightforwardly.
Denoting  $\alpha_{B}  =  \pi \couplingbath^{2} (\partial \rho_{B} (0)) > 0$
(we assumed that $\rho_{B}$ is slowly varying on the frequencies scales at which the  spin
precesses to expand $\rho_{B}$ linearly), this stochastic differential equation is 
(at dominant order $\Sm=S_0$)  our central result Eq.(\ref{eq:rocks}).
Note that we have neglected corrections to $R (u)$
of order $\orderDEUX$.

The electrical current flowing through the system (from lead $a$ to the dot) can also be calculated within the continuous
limit from Eq. (\ref{eq:current1}) : 
\begin{multline}
  \label{eq:current_final}
  I_a = e\int_{-1}^{1}du \frac{\rho(u)}{2\Gamma}  
\sum_{\sigma \sigma'}\sum_{b\ne a} \frac{(1 + \eta \sigma u)(1 + \eta \sigma' u)}{1 + \eta p u} 
\Gamma_a^\sigma\Gamma_b^{\sigma'}  
\times
 \\
\Bigl(n_{a}(x_\sigma) - n_{b} (x_\sigma')\Bigr)
\end{multline}
At a stable fixed point $u$, the density matrix is a Dirac peak
and we get :
\begin{multline}
  \label{eq:current}
  I_a = \frac{e}{2\Gamma}  
\sum_{\sigma \sigma'}\sum_{b\ne a} \frac{(1 + \eta \sigma u)(1 + \eta \sigma' u)}{1 + \eta p u} 
\Gamma_a^\sigma\Gamma_b^{\sigma'}  
\times
 \\ \left[(n(\epsilon -\kappa\sigma u -h \frac{\sigma}{2}-\mu_a) - 
n (\epsilon -\kappa\sigma u -h \frac{\sigma'}{2}-\mu_b)\right]
\end{multline}

At this point let us rediscuss the range of validity of the \nanoLLG equation.
In the derivation given in section \ref{sec:QME} of the master equation, we have used the
condition that the tunneling rates are small compared to the excitation energies
of the grain (to neglect the off-diagonal terms of the density matrix). For the grain,
this condition translates into $\kappa\gg\Gamma$. The derivation was general, and did not
make use of the particular structure of the grain's Hamiltonian. In particular, as we have seen,
the characteristic time scale of the charge and spin dynamics differ by a factor $1/S_0$. This
extends the validity of the \nanoLLG equation to $\kappa\gg\Gamma/S_0$. Physically this can
be seen as follows: as the spin $S_0$ is large and a tunneling electron can only change it
by $1/2$, the time needed for $S_z$ to change significantly is of order $S_0/\Gamma$ while the time
for the internal precession of the magnetization is of order $1/\kappa$. Hence,
when $\kappa\gg\Gamma/S_0$, the system has made many precessions before the spin had the time to change significantly and the dynamics of transverse magnetization is therefore averaged out. 
However, it would be necessary to take the transverse dynamics into account to compute the
$1/S_0$ corrections to the electrical current Eq.(\ref{eq:current}). 

The noise term in Eq.(\ref{eq:rocks}) is of order $1/S_0$ smaller than the main deterministic
term and do not play a major role in general. There are regimes however (when the bias voltage is
high enough) where the deterministic term completely vanishes and the dynamics of the
magnetization becomes completely diffusive. Such a regime has been studied in detailed in Ref.~\cite{waintal2003}.
In order to make contact with this work, let us note that there are sub-dominant
order corrections to $R (u)$, which we have neglected so far.
When $u\rightarrow\pm 1$ however, $R(u)$ vanishes and the terms
of order $\Gamma/\Sm^2$ become important. To make contact with
the results of \cite{waintal2003}, we derive the
evolution of the spin close to the ``pole'' $u=1$.
The correction $\delta R(u)$ of order $\Gamma/S_m^2$ to $R (u)$ reads :
\begin{equation}\label{CorrectionRu}
\delta R (u)  =  \frac{1}{\Gamma \Sm^{2}}
\parent{
\frac{ \check a b - a\check b + u (\check b b -\check a a )}{1  + p\eta u}
+  (1-u^{2} ) f (u)
}
\end{equation}
where $f$ is a regular function in $u=\pm 1$.
For $u$ close to 1, we linearize with $u = 1- {\cal M}/S$ and look for
the (linear) equation of evolution of ${\cal  M}$.
In order to obtain an equation for the (noise-)average $\moy{{\cal
M}}$, we take advantage of the Ito form of the Fokker-Planck equation
and the associated Langevin equation (\ref{eq:rocks}), which implies
$\moy{D (u) \xi_{t}} =0$. Putting together the contributions coming from
Eq.(\ref{eq:rocks}) and Eq.(\ref{CorrectionRu}), we obtain (at the order
$\Gamma/\Sm^2$),
\begin{subequations}
  \begin{align}
\frac{d}{dt} \moy{{\cal  M}} &= \gamma_{+} - (\gamma_{-}- \gamma_{+}) \moy{{\cal M}}
\\
\gamma_{-} &\equiv
\frac{
\sum_{a}
   n_{a} (x_{\uparrow}) \Gamma^{\uparrow}_{a}
  \sum_{b} \check n_{b} (x_{\downarrow})\Gamma^{\downarrow}_{b}
}{\Gamma (1+p\eta )}
\\
\gamma_{+} &\equiv
\frac{
\sum_{a}
  \check  n_{a} (x_{\uparrow}) \Gamma^{\uparrow}_{a}
  \sum_{b} n_{b} (x_{\downarrow})\Gamma^{\downarrow}_{b}
}{\Gamma(1+p\eta )}
\end{align}
\end{subequations}
which is precisely the equation obtained in \cite{waintal2003}.

\section{Phase diagram}
\label{Sec:NanoTorquePhaseDiag}
We now turn to the study of the phase diagram of the  \nanoLLG equation (\ref{eq:rocks}).
We restrict ourselves to the deterministic part and drop the noise part.
As we shall see, the \nanoLLG equation has a structure which is similar to the Landau-Lifshitz-Gilbert (LLG) equation,
which is the fundamental microscopic equation that has been used widely to describe the magnetization
dynamics of system smaller then the exchange length (i.e all excitations are uniform in space, hence the 
system that can be assimilated to a macrospin).
Therefore a joint study gives strong insights on the role of transport in the magnetization dynamics and we start this
section with the LLG equation.
It was understood in the last few years that the non-conservation of the spin current
could lead to a current induced torque~\cite{slonczewski1999}, magnetic relaxation~\cite{tserkovnyak2002}
and interlayer exchange interaction~\cite{slonczewski1989}.
In the framework of the \nanoLLG equation, we will find that the first two physical effects 
can be viewed as different manifestations of one single term in the equation. The latter effect however
is destroyed by the absence of coherence ($T\gg\Gamma$) in our system.  

\subsubsection{The Landau-Lifshitz-Gilbert equation with spin torque}
\label{sec:llg}
The LLG  equation consists of a conservative term (the magnetization precesses around
trajectories of constant energy) plus a phenomenological damping term that allows the system
to relax to its equilibrium. Recently, the LLG equation has also been used to describe the
dynamics of a thin magnetic layer in presence of spin polarized current ~\cite{slonczewski1996}.
In those systems, the polarized current exerts a torque that drives the system out of equilibrium.
The spin torque can be incorporated into the LLG equation through an additional term.

We start by writing the LLG equation for the magnetization of the nanomagnet in presence of the
current induced spin torque, as introduced by Slonczewski \cite{slonczewski1996}. Here
we suppose the magnetic grain is connected to two leads, one of which being magnetic. See
Fig.~\ref{fig:system} for a cartoon of the system.
We denote by $\vec m$  a unit vector pointing in the
direction of the nanomagnets magnetization.
\be
\label{eq:LLG}
\frac{\partial\vec m}{\partial t} = \vec m\times\left[
\frac{1}{\hbar S_0} \frac{\partial E}{\partial \vec m}
+ \alpha \frac{\partial\vec m}{\partial t}
+ \frac{I}{e S_0} g(\vec m\cdot\vec z)\  \vec z\times \vec m  \right]
\ee
where $e$ the charge of the electron,
$I$ the current flowing through the nanoparticle and $S_0$ the total spin of the nanoparticle.
The magnetic energy $E(\vec m)$
consists of a uniaxial anisotropy (along the $z$-axis) and
a Zeeman term that couples to an external magnetic field $\vec H$ (which also lies along the $z$
direction).
\be
E(\vec m)=- \kappa S_0 (\vec m\cdot\vec z)^2 - g \mu_B S_0 \vec H\cdot\vec m
\ee
The terms proportional to the current $I$ corresponds to the spin torque as calculated in
Ref.\cite{slonczewski1996}. Here again we have supposed that the magnetization $\vec m_1$ of the
polarizing lead lies along the $z$-axis. The torque is asymmetric with respect to the parallel/
anti-parallel configuration (with respect to $\vec m_1$) and with good approximation,
the torque is modulated by a function $g(\vec M\cdot\vec z)$ that takes the following form.
\be
g(\vec M\cdot\vec z)= \frac{a}{1+\bar q \vec M\cdot\vec z}
\ee
with $0\le \bar q < 1$. The system having a cylindrical symmetry around the $z$ axis,
Eq.(\ref{eq:LLG}) can be rewritten in spherical coordinate $(\theta,\phi)$ (see Fig.~\ref{fig:system})
and we get,
\be
\left\{
\begin{array}{l}
\dot\theta =    -\alpha \dot\phi \sin\theta -\frac{I}{e S_0} g(\cos\theta ) \sin\theta
\\
\dot\phi \sin\theta = \alpha\dot\theta +\frac{2\kappa}{\hbar} \cos\theta\sin\theta +\gamma H\sin\theta
\end{array}
\right.
\ee
The angle $\phi$ can be eliminated and we get a closed equation for $u=\cos\theta$,
\be
\label{eq:llg-z}
\dot u = \frac{1}{1+\alpha^2} (1-u^2) \left[\alpha \gamma H +\alpha \frac{2\kappa}{\hbar} u +
\frac{I}{e S_0}\frac{a}{1+\bar q u}  \right]
\ee
where $\gamma=g \mu_B/\hbar$ is the gyromagnetic ratio.
Eq.(\ref{eq:llg-z}) is the effective LLG equation along the $z$ axis.
Introducing 
the dimensionless quantities
\begin{align}\label{defredhi}
\bar H &\equiv \frac{g\mu_B H}{2\kappa}
\\
\bar I&\equiv \frac{\hbar a I}{2 \alpha  \kappa e S_0}
\end{align}
we plot its phase diagram  in  Fig.~\ref{fig:phasediagram}.
This diagram was already obtained in  Ref~\cite{bazaliy2004} with a different approach. Here, 
we sketch its construction as it will be useful in the study of the 
\nanoLLG equation.

The phase diagram is constructed from the analysis of the stability of the various fixed points of the equation (for $\dot u=R(u)$, stable fixed points are given by $R(u)=0$ and $\partial_u R(u)<0$
see Fig.\ref{fig:stabilityLLG}): 
the right hand side of Eq.(\ref{eq:llg-z}) vanishes for $u=\pm 1$ and possibly for
up to two other fixed points, depending on $\bar I$ and $\bar H$.
The phase diagram is then determined by 3 curves, which defines 5 regions (See Fig.~\ref{fig:phasediagram}) : 
\begin{itemize}
\item Stability of $u=+1$ :
\be
\bar H +1+\frac{\bar I}{1+\bar q}>0
\ee
\item Stability of $u=-1$:  
\be
\bar H - 1 +\frac{\bar I}{1-\bar q}<0
\ee
\item Existence of two other fixed points $u^*,u^{**}$
\be
(\bar q\bar H -1)^2 - 4 \bar q \bar I >0
\ee 
\end{itemize}
In the following, we will label a region of the parameter space by the list of its {\it stable} fixed points : 
for example the $(-1)$ phase corresponds to one stable fixed point at $u=-1$, $(SP,+1)$ corresponds to two stable
fixed points at $u=-1$ and an intermediate value $-1<u^*<+1$.
We observe that : {\it i)} stable and unstable fixed points alternate between -1 and 1, {\it ii)} if  $u^*<u^{**}$ exists and are located 
between $-1$ and $1$, $u^*$ is stable and $u^{**}$ is unstable (because of the sign of the right hand side of
(\ref{eq:llg-z})). From this remarks and the stability of $u=\pm 1$, we find the regions $(+1)$,$(-1)$,$(-1,+1)$ and $(SP)$ of the phase diagram.
The region delimited by the 2 lines and the parabola (shaded on Fig. \ref{fig:phasediagram}) requires a more detailed
 analysis : on the parabola, 
we have $u^*=u^{**} = -(\bar H + 1/{\bar q})/2$ (they reach $\pm 1$ at the points $A$ and $B$, where the stability lines of -1 and +1 are
tangent to the parabola). A straightforward calculation 
shows that the two others frontiers of this region correspond to $u^* =-1$ and $u^{**}=1$ respectively. Hence, in this region,
$-1<u^*,u^{**}<1$, and we have an intermediate stable SP fixed point (and an intermediate unstable one) : it is the (SP,+1) region.
The five possible fixed points configurations are shown in
Fig.~\ref{fig:stabilityLLG} where a sketch of $\dot u$ is presented.
The presence of the (SP,+1) region is
physically important since it removes the possibility of switching the grain's magnetization
back and forth without hysteresis. We also note that when the current induced torque is
symmetric with respect to the parallel/anti-parallel configuration ($\bar q=0$) the region (SP) and (SP,+1)
are sent to infinity.
\begin{figure}
\vglue +0.45cm
\includegraphics[width=9cm]{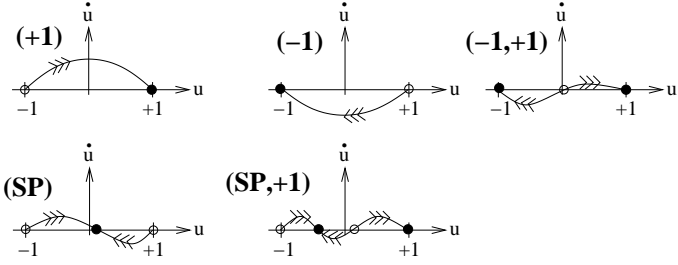}
\caption{\label{fig:stabilityLLG}
Sketch of the different possibilities for the $\dot u$ curves (Eq.(\ref{eq:llg-z}))
leading to the different parts of the phase diagram.
The full circles indicate the stable fixed points while
the empty circles the unstable ones. In diagram $(SP)$ and $(SP,+1)$, an intermediate point $-1<u^*<1$
is stable corresponding to a spin precession state.}
\end{figure}
\begin{figure}
\vglue +0.45cm
\includegraphics[width=9cm]{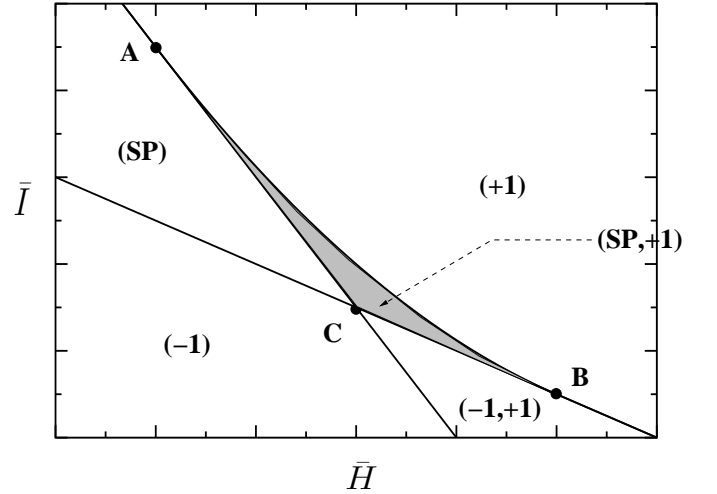}
\caption{\label{fig:phasediagram}
Phase diagram of the LLG equation (\ref{eq:LLG}). The various phases are denoted by
the list of their stable fixed points (See text).
The intersection points have the following coordinates in the $(\bar H, \bar I)$ plane : 
$A =
\left(
-2-\frac{1}{\bar q}  , \frac{(1+\bar q)^2}{\bar q}
\right)$,
$B =
\left(
2-\frac{1}{\bar q} , \frac{(1-\bar q)^2}{\bar q}
\right)$, 
$ C= \left(
-\frac{1}{\bar q}, \frac{1-\bar q^2}{\bar q}
\right) $. 
}
\end{figure}

\subsubsection{The \nanoLLG equation}
\label{sec:NanoTorquePhaseDiag}

In this section, we study the phase diagram of the \nanoLLG equation (\ref{eq:rocks}) at  zero temperature for various sets of  parameters,
both analytically and from the numerical solution of the fixed points equations and analysis of their stability (See Fig. \ref{fig:nanoLLG}).
\begin{figure}[htb]
\vglue +0.45cm
\includegraphics[width=9cm]{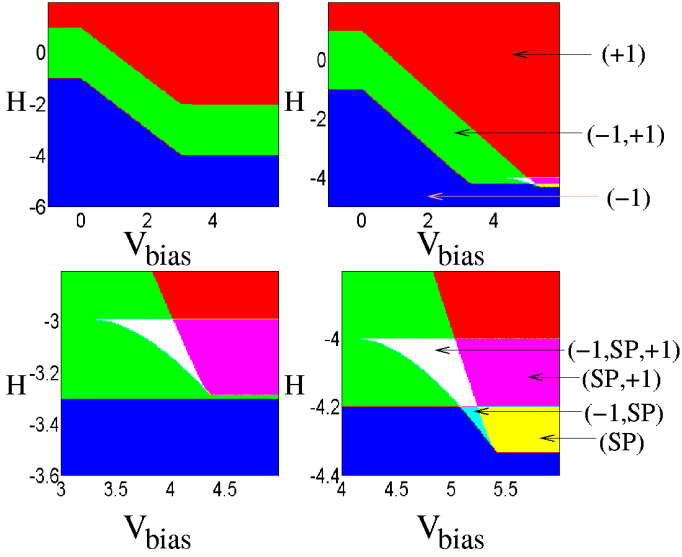}
\caption{ \label{fig:nanoLLG} (Color Online).
Phase diagram of the \nanoLLG equation as a function of  the reduced bias voltage $\bar v$ and
magnetic field $\bar h$ ({\it i.e.} measured respectively in unit of $\kappa/e$ and $2\kappa/\gamma$). The tunneling rates
are in unit of $8 S_0 \alpha_B \kappa/\hbar$ and  $V_{\rm bias}$ has been translated of the charging energy.
The color code corresponds to the stable fixed points as indicated on the figure. SP,+1 and -1 indicate the corresponding stable fixed points $u=\pm 1$ or $u^*\ne \pm 1$ for the Spin Precession (SP) state.
Upper Left panel: $\Gamma^\uparrow_{\rm left}=4.5$, $\Gamma^\downarrow_{\rm left}=1.5$, $\Gamma^\uparrow_{\rm right}=1.5$ and
$\Gamma^\downarrow_{\rm right}=4.5$. 
Upper Right panel: $\Gamma^\uparrow_{\rm left}=12$, $\Gamma^\downarrow_{\rm left}=4$, $\Gamma^\uparrow_{\rm right}=8$
and $\Gamma^\downarrow_{\rm right}=8$.
Lower Right panel: zoom of upper right panel. 
Lower Left panel: $\Gamma^\uparrow_{\rm left}=8$. $\Gamma^\downarrow_{\rm left}=2$, $\Gamma^\uparrow_{\rm right}=5$
and $\Gamma^\downarrow_{\rm right}=5$. }
\end{figure}
The phase diagram is obtained at zero temperature by analogy with the LLG equation~(\ref{eq:llg-z}) 
with the additional complexity that the 
presence of the Coulomb blockade phenomena makes the ``current'' term in (\ref{eq:rocks}) 
more complicated than the one of the LLG equation. 
Indeed this  ``current'' term can be switched on and off as $u$ 
goes through the critical values where the addition energies $\Delta E^\sigma (u)=\epsilon -\kappa\sigma u -h \frac{\sigma}{2}$ are equal to the chemical potential 
of the leads: the conducting channels for majority (resp. minority)
electrons can open or close when $u$ varies.

The simplest case is the symmetric case, where the torque asymmetry $q$ vanishes.
this removes the possibility of exciting spin precession states (there can only be one intermediate fixed point $u^*$
and according to the sign of the coefficient of $u$ it must be repulsive).
In the upper left panel of Fig.\ref{fig:nanoLLG}, we have represented
such a case, where the two leads are magnetic with same tunneling rates, and lie in the anti-parallel configurations. 
This is the most favorable case for current induced switching: not only up electrons have a higher
probability to enter the grain than the down electrons, but in addition, once in, they have a lower probability 
to get out so that the imbalance between up and down current is enhanced.

To study the more interesting case $q\neq 0$, we will now restrict ourselves  to the case where 
the chemical potential of the second lead is very low. In the Lower Left panel of Figure (\ref{fig:nanoLLG}), 
we display such a case : note the presence of two additional regions of stability 
$(SP,+1)$ and $(-1,SP,+1)$, the latter having no equivalent in the LLG equation.
To gain analytical understanding, it is convenient to rewrite the \nanoLLG equation as : 
\be
\label{eq:rocks2}
\dot{u} = (1-u^{2}) 
\biggl( \bar h +  u  + \frac{\hat A(u)}{ 1 + q u}
\biggr)
\ee
where we rescaled the time by $\bar t  \equiv (2\alpha_B\kappa) t$, restricted ourselves to  $\eta=1$ and defined : 
\begin{subequations}
  \begin{align}
  \label{eq:defhvtilde}
 \bar h  &= \frac{h}{2\kappa}
\\
\bar v &= \frac{\mu_L-\epsilon}{\kappa} 
\\
 \hat A (u)  &= \frac{\hbar }{8\kappa \alpha_B S_0}
 \left( (1-q) \Gamma_L^\uparrow \theta_\uparrow(u)  - (1+q) \Gamma_L^\downarrow \theta_\downarrow(u) \right)
\end{align}
\end{subequations}
where $\theta_\sigma(u) = \theta( \sigma \bar h  + \sigma u + \bar v)$ with $\sigma =\pm 1$ and 
$\theta$ the Heaviside function.
$\theta_\sigma(u)$ is 1 when the channel $\sigma$ is open 
and 0 otherwise, for a given  value of $u$. 
When all channels are open, $\hat A(u)$ takes the value : 
\begin{equation}
  \label{eq:defA}
 \hat A(u) = \Theta  \equiv   \frac{\hbar }{8\kappa \alpha_B S_0}
 \left(
   (1-q) \Gamma_L^\uparrow   - (1+q) \Gamma_L^\downarrow 
 \right)
\\
\end{equation}  
%

We start with the stability analysis of the two fixed points $u=\pm 1$. It is summarized on Fig. \ref{fig:phasediagram.construct} : 
we plotted the various regions where up and down channels are open and closed, together with the line of stability
for (+1) and (-1) in thick solid line.
\begin{figure}
\includegraphics[width=9cm]{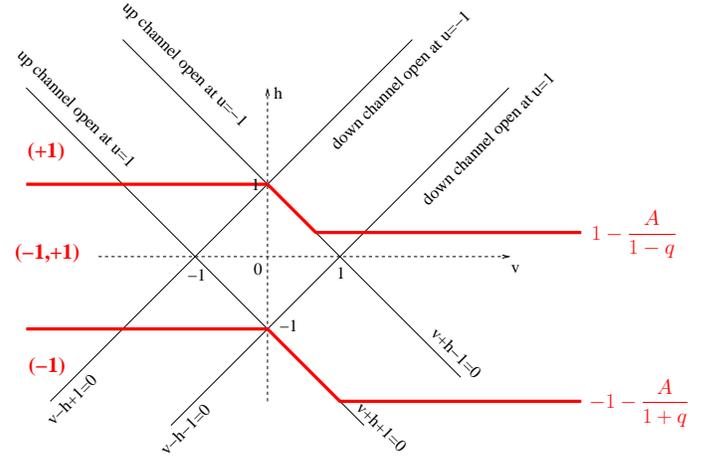}
\caption{\label{fig:phasediagram.construct}
The upper (lower) thick line corresponds to the limit of stability of the $u=-1$ ($u=+1$) fixed point in the
tension bias-magnetic field plane. The various thin lines correspond to the opening of the the spin up/down channel
depending on the magnetic configuration.
}
\end{figure}
The phase diagram is then established in the following way:  first we concentrate of the large bias region 
(the right part of the phase diagram), where all channels are open.
In that case, $ \hat A(u) = \Theta$ so we keep $A$ fixed and vary $q$, in order to use our previous results  on LLG equation;
second we reduce the voltage to see when various fixed points are destabilized when some channels closes.
At high voltage, we find 4 possibilities, depending on how $A$ compares to the $\bar I$-coordinate of points $B$ and $C$ 
on Fig \ref{fig:phasediagram} : 
\begin{itemize}
\item $\Theta\leq \frac{(1-q)^2}{q}$ : this is the small $q$ region. $\Theta$ is lower than the $\bar I$-coordinate of the $B$ point 
and one can not excite $(SP)$ fixed points.
The phase diagram is similar to the Upper Left panel of Fig. \ref{fig:nanoLLG}.
\item $ \frac{(1-q)^2}{q} \leq  \Theta \leq \frac{1-q^2}{q}$ : $\Theta$ is between the $\bar I$-coordinate of the 
$B$ and the $C$ point on Fig. \ref{fig:phasediagram}.
Therefore, for some magnetic fields, there is a stable $(SP)$ point and the phase diagram of the \nanoLLG equation is  similar to
the lower left panel of Fig. \ref{fig:nanoLLG}. If one starts in the $(SP)$ region and lower the bias, the $(SP)$ fixed point will be destabilized 
{\it by the closing of the up channel at $u=u^*$}, not at $u=-1$. This happens at a lower bias, and leads 
to the existence of the $(-1,SP,+1)$ region.
Note that the first inequality is equivalent to requiring that  $\bar h + u + \Theta/(1+q u)$ has a negative slope at $u=-1$.
At the bottom of the  $(-1,SP,+1)$ region, the frontier merges with the stability line of -1, since $u^*=-1$ on $[B,C]$
on Fig. \ref{fig:phasediagram}. On the contrary, this region is limited at higher magnetic field $\bar h$, corresponding the
frontier of region $(SP,+1)$ and $(+1)$ on  Fig. \ref{fig:phasediagram} : at this point, $(q\bar h - 1)^2 - 4q\Theta =0$
hence $\bar h_{max} = \frac{1 - 2 \sqrt{q\Theta}}{q}$; moreover, on this frontier $u^*$ is not $\pm 1$, so the upper limit of the  
 $(-1,SP,+1)$ region does not merge with the stability line of -1.
\item $ \frac{1-q^2}{q} \leq  \Theta  \leq \frac{(1+q)^2}{q} $: In this case, $\Theta$ is bigger than the $\bar I$-coordinate of the $C$ point in on Fig. \ref{fig:phasediagram} but lower than the A coordinate, 
and there is two additional regions (as a function of the magnetic field) : $(SP)$ and $(-1,SP)$.
This case is presented on the Upper Right panel with a zoom in the Lower Right panel of Fig. \ref{fig:nanoLLG}. 
The origin of the $(-1,SP,+1)$ region is analogous to the previous case. 

\item $ \frac{(1+q)^2}{q} \leq  \Theta$. In this regime, $\Theta$ is bigger than the $\bar I$-coordinate of the $A$ point in  Fig. \ref{fig:phasediagram}. 
This regime corresponds to rather high polarization of the leads and is not illustrated in Fig. \ref{fig:nanoLLG}. 

\end{itemize}

The $(-1,SP,+1)$ region has no equivalent in the LLG equation. In this region,
the up spin channel is blocked at $u=-1$ so that the torque term does not destabilize the $(-1)$ phase. For
$u$ slightly higher, this channels opens up, and stabilizes a SP fixed point (provided the torque
is assymetric enough).

The behavior in the regions with multiple fixed points is clearly highly hysteretic, and the hysteretic loops
in magnetic field or bias voltage are very different. 
As an illustration, we present in Fig.~\ref{fig:current}, two cross-sections of the Lower Right panel of
Fig.~\ref{fig:nanoLLG}. In the left panel, we present the current $I$ as a function of $V_{\rm bias}$ with
$H$ chosen such that the system goes through the $(-1,+1)$,  $(-1,SP,+1)$ and $(SP,+1)$ phases. We find two different unconnected
 loops: if the system is initially in $u=+1$, it will stay there, no matter the value 
of $V_{\rm bias}$. On the contrary, if the system is initially in a SP state or in $u=-1$, it will switch, hysteretically
 between the $SP$ and the $-1$ state. In the right panel, $V_{\rm bias}=5.5$ and we plot the current $I$
as a function of magnetic field $H$. In the hysteresis loop, the value of $u^*$ in the $SP$ states evolves 
continuously as a function of $H$ so that the current no longer takes discrete values. When the system
is in a SP state, in addition to the DC current calculated here, small AC corrections are expected.
This AC current, though small ($\sim 1/S_0$ smaller than the main contribution) is strongly picked on
the precession frequency of the SP state ($\propto \kappa u$) and can be used for an experimental
proof of the presence of the spin precessing state. In this article however, we did not calculate
this AC current, which would require to take into account the off-diagonal part of the density matrix. 
\begin{figure}[htb]
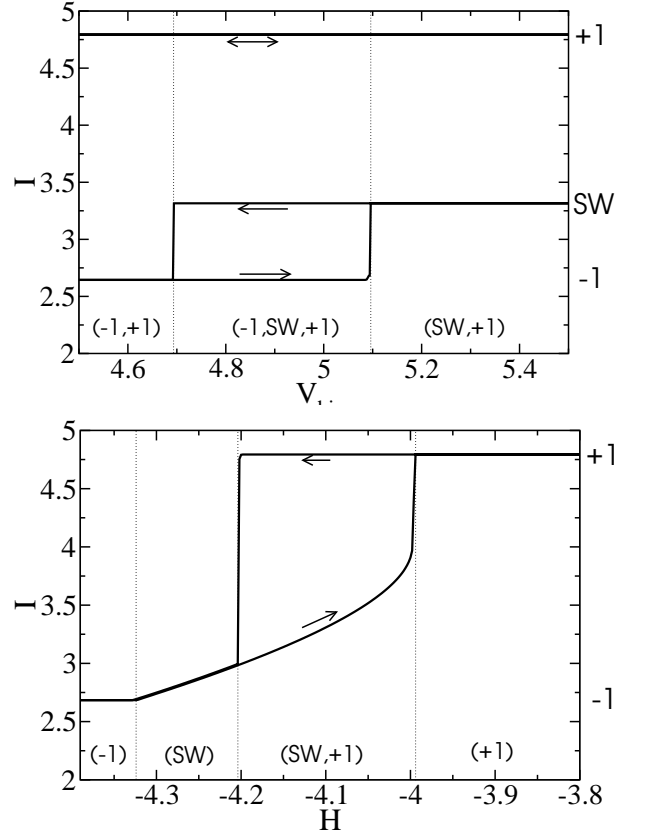

\vglue +0.45cm
\includegraphics[width=8cm]{current-bias.eps}
\includegraphics[width=8cm]{current-h.eps}
\caption{\label{fig:current} Current as a function of bias voltage $V_{\rm bias}$ (left panel) and
magnetic field $H$ (right panel), measured respectively in unit of $\kappa/e$ and $2\kappa/\gamma$.
The voltage $V_{\rm bias}$ is shifted by $\epsilon=20$.
The tunneling rates are in unit of $8 S_0 \alpha_B \kappa/\hbar$ and  $V_{\rm bias}$ has been translated 
of the charging energy. Same system as in the Upper Right panel of Fig.~\ref{fig:nanoLLG}:
$\Gamma^\uparrow_{\rm left}=12$, $\Gamma^\downarrow_{\rm left}=4$, $\Gamma^\uparrow_{\rm right}=8$
and $\Gamma^\downarrow_{\rm right}=8$. In the left panel, $H=-4.1$ and the two curves correspond to two ways
of preparing the system, see text. In the right panel, $V_{\rm bias}=5.5$}
\end{figure}

\subsubsection{\nanoLLG equation without bias voltage}

In addition to the current induced magnetic reversal and spin precession state described in the
previous section, the \nanoLLG equation also describes the extra relaxation induced by the presence of the leads.
It is interesting to note that the current induced torque and the extra relaxation 
are just two manifestations of a single phenomena while in the scattering formalism, they were derived as two
different contributions~\cite{waintal2000,tserkovnyak2002}.
When no bias voltage is put across the sample, the \nanoLLG equation simplifies into
(omitting the noise term)
\begin{multline}
\label{eq:rocks-equilibrium}
\dot{u} = (1-u^{2}) 
\biggl[
 \alpha_{B} (h + 2\kappa u) +
\frac{\Gamma (1-p^2)}{8S_{0} ( 1 + q \eta u)} \times \\
\frac{\sinh((\kappa u+h/2)/kT)}{\cosh(\epsilon/kT) +\cosh((\kappa u+h/2)/kT)}
\biggr]
\end{multline}
In the large temperature limit, this equation takes exactly the form of the LLG
equation~(\ref{eq:llg-z}) with an effective damping constant $\alpha_{\rm eff}$
given by:
\be
\alpha_{\rm eff}= \alpha_{B} + \frac{\Gamma (1-q^2)}{16 kT S_{0}  ( 1 + q \eta u)}
\ee
When the leads are magnetic and $q\neq 0$, their contributions to the damping becomes anisotropic: the
phase diagram is unaffected but the relaxation rates for $u=\pm 1$ differ. In the low temperature limit, the lead-induced relaxation rate saturates to
$\pm\frac{\Gamma (1-q^2)}{8 S_{0}  ( 1 + q \eta u)}$ depending on which (up or down) channels arekil
opened, and vanishes when both channels are opened.

The zero bias equation for the current simplifies into,
\be
  \label{eq:current-equilibrium}
  I_a = \frac{e}{2\Gamma}  
\sum_{b\ne a} \frac{1 - u^2}{1 + \eta q u} 
(\Gamma_a^\uparrow\Gamma_b^{\downarrow} - \Gamma_a^\downarrow\Gamma_b^{\uparrow})  
\Bigl(n(\Delta E^{\uparrow}) - n (\Delta E^{\downarrow})\Bigr)
\ee
Eq.(\ref{eq:current-equilibrium}) vanishes at equilibrium where $u=\pm 1$ as it should. However if the 
system is initially out of equilibrium, a DC current will flow during the relaxation process. If the 
system is maintained in a spin precession state (with radio frequency magnetic field for instance)
a rectified DC current will follow.

\section{Conclusion}

 The discussion of the experimental relevance of the \nanoLLG equation and of
the possibility of observing some of the effects described in this paper was already done
in Ref.\onlinecite{PRLWaintalParcollet2005}. Beside its application to ultrasmall nanomagnets,
Eq.~(\ref{eq:rocks}) is also interesting from the point of view of spintronic as it can give valuable 
insights on various effects usually discussed in more macroscopic systems. 
Hence, we conclude this article by coming back to a few key points raised in this
article and contrasting them with
the usual theoretical treatment done on macroscopic systems.
(i) There is only one kind of electrons in our system, responsible for both the 
magnetization and transport. The possibility of writing a {\it closed} equation for the magnetization
only comes from the separation of time scales between magnetic and charge degree of freedom.
This timescale separation itself comes from the fact that many ($\propto S_0$) tunneling events
are needed to change the magnetization significantly (far from equilibrium) or from
the smallness of the Clebsch-Gordon coefficient (close to equilibrium, see~\cite{waintal2003}).
(ii) One usually distinguish between two kinds of tunneling events, either on a majority
or a minority state. However, in the case studied in this paper, where the system precesses fast around
its easy axis, the good quantization axis for the spin is the easy axis of the grain not the direction
of the magnetization. Hence we have
to distinguish four kind of tunneling events depending on the orbital $\alpha$ (majority/minority)
and the spin on the $z$ axis (up/down). 
(iii) Low energy excitations are easily identified in our system. In particular, we can separate
the magnetization excitations (excitations of type E) from the spin accumulation
(excitations of type C and D, see Fig.\ref{fig:excited}). The latter correspond to electrons
going to/from a majority state from/to a minority one hence changing the total spin $S$ while
the former correspond to a change of $S_z$ only. The \nanoLLG equation has been derived in a regime
where only one one-body state carries the current, hence there is no spin accumulation in our
system. In this case, the spin torque and spin accumulation phenomena completely decouple.
(iv) In contrast, the tunneling magneto-resistance is very sensitive to the presence of 
spin-accumulation which gives rise to non-equilibrium peaks in spectroscopy experiments.
(v) The three (up to now) phenomena associated with the (non conservation of) the spin current in 
magnetic systems are the current induced spin torque, spin pumping induced relaxation and 
magnetic exchange interaction. It is remarkable that the first two are contained by a single
term in the \nanoLLG equation. The exchange interaction however, which relies on equilibrium
 spin current and on spin coherence is not captured by the master equation.
(vi) The possibility of stabilizing spin precession states have a similar origin as in 
nanopillars\cite{bazaliy2004}: the amplitude  of the torque decreases when the system becomes out of equilibrium.
The phase diagram is however more complex as the different channels for spin up and down can be
opened or closed. (vii) The torque term in nanopillars vanishes as the sinus of the angle between
the polarizing layer and the thin free layer magnetizations. In contrast, the torque term in our nanomagnet can
exist even in the {\it collinear} regime and the torque vanishes as the sinus of the angle between the
easy axis and the grain magnetization. Hence, we expect drastic changes in our system behaviour when the temperature
becomes lower than the tunneling rates and we enter the coherent tunneling regime.
 

\acknowledgments
We thank F. Portier for useful comments.

\appendix


\section{Diagonal approximation}\label{diagonal_approx}

In this appendix, we examine under which conditions the solution of an equation of the
form

\be
\label{eq:osc}
\dot{X} = Q X + \sum_{n=0}^{N} M_{n} e^{i \omega_{n} t} X
\ee
converges to the solution of the equation
\[
\dot{X} = Q  X
\]
in the limit $\omega_{i}\rightarrow \infty$,  where $X$ is a complex vector  and $M_{n}$ , $Q$ are $n\times n$ matrices.
Eq.(\ref{Kubo.proj.z2}) for the density matrix is a special case of equation (\ref{eq:osc}).
In analogy with the scalar case, we define
\[
A_i \equiv  \exp\left[\frac{e^{i\omega_{i} t}}{i\omega_{i}} M_i\right]
\qquad
Y= \prod_{i=0}^N A_i^{-1} X
\]
and arrive at
\def\om{\{\omega_i\}}
$$
\dot Y_{\om}(t) = M_{\om} (t) Y_{\om}(t)
$$
where $M_{\om} (t)\rightarrow Q$ {\it uniformly in } $t$ for
$\om \rightarrow \infty$.
Denoting by $Y_{\infty}$ the solution of $\dot{Y} = Q Y$,
we then have $Y_{\om} (t) \rightarrow Y_{\infty} (t)$ for all $t$.
However, it does not imply that
$\lim_{t\rightarrow \infty}{Y_{\infty} (t)}
= \lim_{\om \rightarrow \infty}\lim_{t\rightarrow
\infty}Y_{\om} (t)
$.
A trivial example where these two limits cannot be exchanged is
the (scalar) equation $\dot y = -a y$. There, $\lim_{a\rightarrow 0} \lim_{t\rightarrow\infty}
y_a(t) = 0$ while  $\lim_{t\rightarrow\infty} \lim_{a\rightarrow 0} y_a(t) = 1$.
Coming back to the $Q$ matrix, we cannot interchange the two limits whenever $Q$
has vanishing eigenvalues. In the case we are interested in, the matrix $Q$ 
describes a Markov process and hence has all its eigenvalues strictly negative\cite{doob1953} 
but one corresponding to the steady state. As the Markov process 
conserves the trace of the density
matrix (for all $\om$), we can project the dynamics into the corresponding hyperplane
where all the eigenvalues of $Q$ are strictly negative and therefore the two limits can
safely be interchanged. We conclude that when $\om\rightarrow\infty$, one can neglect the
oscillating terms.


\newcommand{{{\PRB}}}{{{Phys. Rev. B}}}\newcommand{{{\PRL}}}{{{Phys. Rev. Lett}}}\newcommand{{{\NPB}}}{{{Nucl. Phys.}}}\newcommand{{{\RMP}}}{{{Rev. Mod. Phys.}}}\newcommand{{{\ADV}}}{{{Adv. Phys.}}}

\end{document}